\documentclass{article}
\usepackage{ijcai23}

\usepackage{times}
\usepackage{soul}
\usepackage{url}
\usepackage[hidelinks]{hyperref}
\usepackage[utf8]{inputenc}
\usepackage[small]{caption}
\usepackage{graphicx}
\usepackage{amsmath}
\usepackage{amsthm}
\usepackage{booktabs}
\usepackage{algorithm}
\usepackage{algorithmic}
\usepackage[switch]{lineno}

\usepackage{amssymb}
\usepackage{bm}
\usepackage[table,xcdraw]{xcolor}
\usepackage[normalem]{ulem}
\usepackage{subfigure}
\usepackage{multirow}
\usepackage{newfloat}
\usepackage{listings}

\useunder{\uline}{\ul}{}
\theoremstyle{definition}
\newtheorem{defn}{Definition}
\newcommand{\mb}{\mathbf}

\title{VR-GNN: Variational Relation Vector Graph Neural Network for Modeling both Homophily and Heterophily}
\author{
  Fengzhao Shi$^{1,2}$\thanks{Equal contribution}\and
  Ren Li$^{1,2}$\footnotemark[1]\and
  Yanan Cao$^{1,2}$\thanks{Corresponding author}\and
  Yanmin Shang$^{1,2}$\and
  Lanxue Zhang$^{1,2}$\and\\
  Chuan Zhou$^{3}$\and
  Jia Wu$^{4}$\And
  Shirui Pan$^{5}$
\affiliations
$^1$Institute of Information Engineering, Chinese Academy of Sciences\\
$^2$School of Cyber Security, University of Chinese Academy of Sciences\\
$^3$Academy of Mathematics and Systems Science, Chinese Academy of Sciences\\
$^4$Macquarie University\\
$^5$Griffith University\\
\emails
\{shifengzhao, liren, caoyanan, shangyanmin, zhanglanxue\}@iie.ac.cn,\\
zhouchuan@amss.ac.cn, jia.wu@mq.edu.au, s.pan@griffith.edu.au
}

\begin{document}

\maketitle

\begin{abstract}
  Graph Neural Networks (GNNs) have achieved remarkable success in diverse real-world applications. Traditional GNNs are designed based on homophily, which leads to poor performance under heterophily scenarios. 
  Current solutions deal with heterophily mainly by mixing high-order neighbors or passing signed messages. However, mixing high-order neighbors destroys the original graph structure and passing signed messages utilizes an inflexible message-passing mechanism, which is prone to producing unsatisfactory effects.
  To overcome the above problems, we propose a novel GNN model based on relation vector translation named \textbf{V}ariational \textbf{R}elation Vector \textbf{G}raph \textbf{N}eural \textbf{N}etwork (\textbf{VR-GNN}). 
  VR-GNN models relation generation and graph aggregation into an end-to-end model based on Variational Auto-Encoder. The encoder utilizes the structure, feature and label to generate a proper relation vector for each edge. 
  The decoder achieves superior node representation by incorporating the relation vectors into the message-passing framework. 
  VR-GNN can fully capture the homophily and heterophily between nodes due to the great flexibility of relation translation in modeling neighbor relationships.
  We conduct extensive experiments on eight real-world datasets with different homophily-heterophily properties and verify the effectiveness of our method. 
\end{abstract}
\section{Introduction}
\begin{figure}[ht]
	\centering
	\subfigure[Passing Signed Message]{
		\includegraphics[width=0.47 \columnwidth]{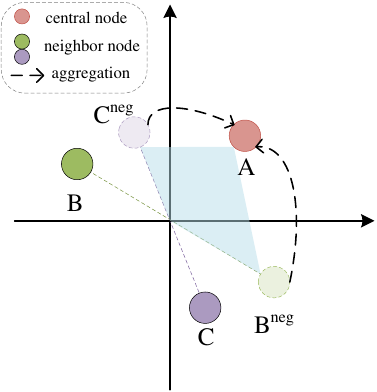}
	}
	\subfigure[Relation Vector Translation]{
		\includegraphics[width=0.47 \columnwidth]{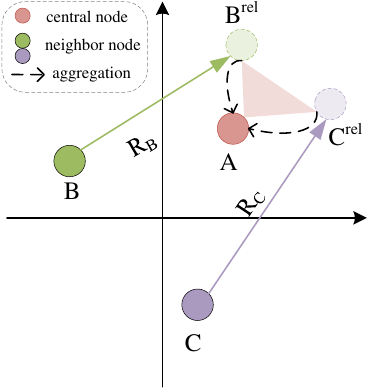}
	}
	\caption{Passing signed message and relation vector translation. The nodes $\rm{B}$ and $\rm{C}$ are the neighbors of $\rm{A}$. Different colors of nodes represent different classes. The $\rm{B}^{neg}$ (or $\rm{C}^{neg}$) and $\rm{B}^{rel}$ (or $\rm{C}^{rel}$) represent the negative message and message transformed by relation vector of node $\rm{B}$ (or $\rm{C}$) respectively. $\rm{R}_B$ (or $\rm{R}_C$) represents the relation vector of node $\rm{B}$ (or $\rm{C}$). Dash lines with arrows represent aggregation operation.}
	\label{fig:motive}
\end{figure}
Graph Neural Networks (GNNs) have revealed superior performance on various real-world graph  applications, ranging from social networks \cite{social2}, citation networks \cite{gat} to biological networks \cite{protein}.
Graph Convolutional Network (GCN) \cite{gcn} and its variants \cite{appnp,gcnii} learn node representations via smoothing the features between neighbor nodes. 
The smoothing operation is suitable for homophilic graphs \cite{homobook}, where connected nodes tend to possess similar features and belong to the same class. 
However, on heterophilic graphs \cite{hete1} where connected nodes have dissimilar features and different labels, the traditional GCNs suffer from poor performance and even underperformance Multi-layer Perceptron (MLP) that completely ignores the graph structure \cite{deeper}.

Recently several efforts have been proposed to achieve heterophily-based GNNs. The algorithms can be mainly divided into two families based on designing methodologies: mixing high-order neighbors \cite{geom,glognn} and passing signed messages \cite{fagcn,gprgnn}.
The approach of mixing high-order neighbors expects to aggregate more homophilic nodes and remove heterophilic nodes. However, its performance is limited by the structure extracted and priors used \cite{wrgat}, which makes it more likely to produce a loss of information compared to directly modeling original graph \cite{hete_book}.
The approach of passing signed messages uses positive and negative signs to modify neighbor information. 
Under this type of method, the neighbors of different classes send negative messages to each other for dissimilating their features, and those of the same class pass positive messages for assimilation.
However, the single numerical sign suffers from limited expressing capacity, which causes the modeling to be inflexible and insufficient. 
As shown in figure \ref{fig:motive} (a), the heterophilic neighbors $\rm{B}$ and $\rm{C}$ deliver their negative messages $\rm{B}^{neg}$ and $\rm{C}^{neg}$ to $\rm{A}$. 
However, the scalar weight (normally within $[-1, 1]$ to maintain numerical stability during message passing \cite{fagcn}) restricts $\rm{B}^{neg}$/$\rm{C}^{neg}$ to moving along the dotted line, hence the new aggregated representation of $\rm{A}$ is confined to the blue shaded part, which does not have the desired effect on extending distance between heterophilic neighbors.

To overcome the above problems, we introduce relation vectors into the message-passing framework. 
Inspired by the relation idea of knowledge graphs (KGs) \cite{kg}, we consider the connections also serve as the relationship between nodes. 
The message passing between nodes could be described by the addition translation of the relation vector, which is similar to the form of TransE, a classical translation model for KG embedding \cite{transe}.
The demonstration is shown in figure \ref{fig:motive} (b), neighbors $\rm{B}$ and $\rm{C}$ translate their features to $\mathrm{B}^{rel}$ and $\mathrm{C}^{rel}$ by the relation vector $\rm{R}_B$ and $\rm{R}_C$. 
Then after aggregating, central node $\rm{A}$ could obtain a more preferred representation in the red area to dissimilate with $\rm{B}$ and $\rm{C}$. 
Compared with signed scalar weights, relation vectors are more flexible and expressive for modeling homophily and heterophily between nodes, which helps to achieve more adaptive message-passing algorithm. 

Based on the above idea, we present a novel method named \textbf{V}ariational \textbf{R}elation Vector \textbf{G}raph \textbf{N}eural \textbf{N}etwork (\textbf{VR-GNN} for short). 
VR-GNN builds the framework based on Variational Auto-Encoder (VAE) \cite{vae}.
The encoder of VR-GNN treats relation vectors as hidden variables and adopts variational inference to generate it based on the graph structure, feature and label, which involve homophily and heterophily at different aspects \cite{wrgat,DMP,cpgnn}. 
The decoder incorporates generated relation vectors into the message-passing mechanism, where messages are computed by translating neighbors along connections. 
Because relation vectors have encoded the homophily/heterophily property of each edge, the translation could produce a suitable assimilation/dissimilation effect between neighbors.
Finally, the model takes the output node representation to perform the downstream classification task and achieves SOTA performance on eight various datasets.

In summary, our main contributions are as follows:
\begin{itemize}
  \item We propose a new message passing mechanism grounded on relation vector translation, for modeling both homophilic and heterophilic connections in graphs.
  Compared  to previous approach of singed message passing, relation vectors are more flexible and possess a more expressive capacity. 
  \item We propose a novel method named \textbf{V}ariational \textbf{R}elation Vector \textbf{G}raph \textbf{N}eural \textbf{N}etwork (\textbf{VR-GNN}). VR-GNN builds the framework based on Variational Auto-Encoder (VAE), and provides an effective end-to-end solution for both relation vector generation and relation guided message passing.
  \item Through extensive experiments on eight common homophilic and heterophilic datasets, we demonstrate the validity of our introduced relation vector concept and VR-GNN method.
\end{itemize}

\section{Preliminary}

\subsection{Problem Definition}
A graph can be denoted as $\mathcal{G}=(\mathcal{V}, \mathcal{E})$ with a set of nodes $\mathcal{V} = \{v_1, v_2, \cdots, v_N\}$ and a set of edges or connections $\mathcal{E}$. The connections of a graph can be described by its adjacency matrix $\mathbf{A}=[a_{ij}] \in \{0, 1\}^{N\times N}$, where $N = |\mathcal{V}|$ is the number of nodes, and $a_{ij}=1$ means node $v_i$ and $v_j$ has a connection $e_{ij}$ between them. 
The node feature matrix of a graph can be denoted as $\mathbf{X} \in \mathbb{R}^{N \times F}$, where $F$ is the feature dimension per node. $\mathbf{x}_i \in \mathbb{R}^F$ denotes the $i$-th row of $\mathbf{X}$ and corresponds to the feature of node $v_i$. 
In this paper, we focus on the semi-supervised node classification task, which aims to learn a mapping $f: \mathcal{V} \to \mathcal{C}$, where $\mathcal{C}=\{c_1, c_2, \cdots, c_M\}$ is the label set with $M$ classes, given $\mathbf{A}$, $\mathbf{X}$ and partially labeled nodes $\{(v_1, y_1), (v_2, y_2), \dots\}$ with $y_i \in \mathcal{C}$. 

\subsection{Message-passing Framework}
Currently, most graph neural networks (GNNs) apply message-passing framework \cite{messagepassing} to formulate their workflow. Normally the message-passing framework consists of $L$ layers, and in each layer $l$, it aggregates neighbor and central nodes with the rule of:
\begin{equation}
	\begin{split}
		\mb{h}_{i}^{l+1} &= \mathrm{U}(\mb{h}_{i}^{l}, \mathrm{AGG}(\mb{h}_{j}^{l}: j \in \mathcal{N}_{i}))
	\end{split}
\end{equation}
where $\mb{h}_{i}^{l}$ denotes the embedding of node $v_i$ in layer $l$; $\mathcal{N}_{i}$ is the neighbor set of node $v_i$; $\mathrm{AGG}(\cdot)$ is the neighbor aggregation function; $\mathrm{U}(\cdot)$ is the updating function, to renew the central node embedding with aggregation information and original node information.

\subsection{Homophily Ratio}
Here we introduce the concept of homophily ratio \cite{geom} to estimate the homophily level of a graph.
\begin{defn}[Homophily ratio]
  The homophily ratio of a node $v_i$ is the proportion of its neighbors belonging to the same class as it. The homophily ratio of a graph is the mean of homophily ratios of all its nodes:
  \begin{equation*}
    \mathcal{H}=\sum_{v_i\in \mathcal{V}}\frac{|\{e_{ij}:e_{ij}\in \mathcal{E} \wedge y_{i}=y_{j}\}|}{|\mathcal{N}_{i}|} \in [0,\, 1]
  \end{equation*}
\end{defn}
A high homophily ratio represents the graph possesses a strong homophily property, and a low homophily ratio indicates a weak homophily property or strong heterophily property. 

\section{Methodology}

\begin{figure*}[ht]
	\centering
	\includegraphics[width=0.9\linewidth]{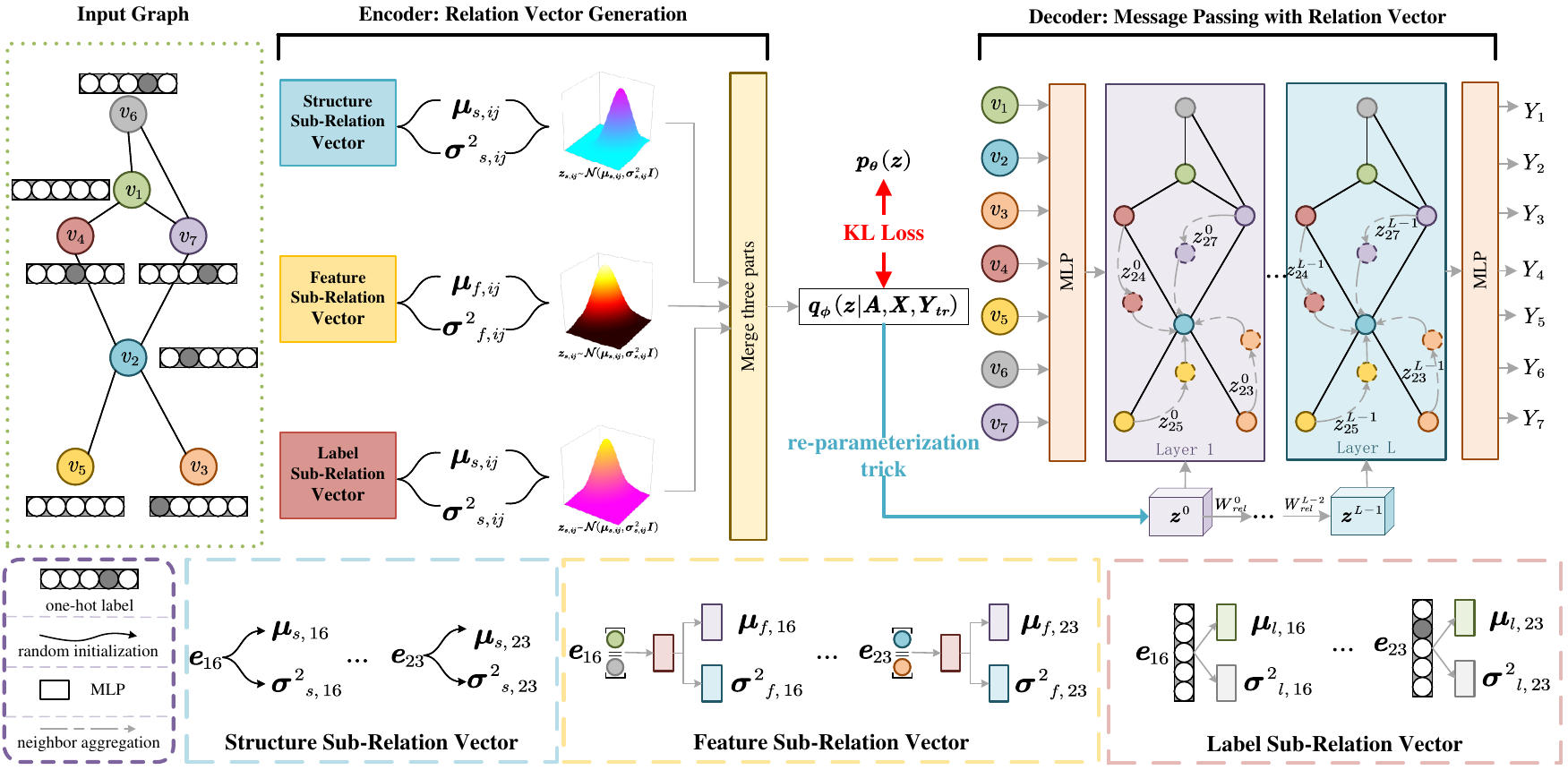}
	\caption{The architecture of VR-GNN. It consists of two components: an encoder to generate relation vectors by combining structure, feature and label information, and a decoder to achieve node classification with generated relation vectors. 
 The encoder generates three types of sub-relation with variational inference and composes them into final relation vectors. The three sub-relations are set as multivariate normal distribution with mutual independence. The decoder utilizes the relation vectors to translate original neighbor features into proper messages and aggregates them for the node representations.}
	\label{fig:model}
\end{figure*}

\subsection{VR-GNN Framework}
The core idea of VR-GNN is to introduce relation vectors to describe diverse homophilic and heterophilic connections of the graph, for helping GNN achieve a more effective message passing. 
Furthermore, we treat such process as an encoder-decoder paradigm, to firstly encode the connection characteristics into relation vectors, then decode the relation vectors through GNN to complete the downstream task, i.e. node classification here. 

In this work, we take Variational Auto-Encoder (VAE), a popular probabilistic technique to encode/decode hidden embedding of the data \cite{vae}, as our overall framework. Specifically, we treat the relation vector of graph connections as latent variable $\mathbf{z}$, and the node classification as a prediction process guided by $\mathbf{z}$, hence the process of VR-GNN can be formularized as following:
\begin{equation}
  \begin{split}
    & p_{\theta}(\mathbf{Y} | \mathbf{A}, \mathbf{X}, \mathbf{Y}_{tr}) \\
    & = \int p_{\theta}(\mathbf{z}| \mathbf{A}, \mathbf{X}, \mathbf{Y}_{tr}) \, p_{\theta}(\mathbf{Y} | \mathbf{z}, \mathbf{A}, \mathbf{X}, \mathbf{Y}_{tr}) \, \mathrm{d} \mathbf{z}
  \end{split}
  \label{eq: e1}
\end{equation}
where $\mathbf{A}$ is adjacency matrix; $\mathbf{X}$ is node feature matrix; $\mathbf{Y} \in \mathbb{R}^{N \times M}$ is label matrix; $\mathbf{Y}_{tr}$ is training label matrix; $\theta$ denotes learnable parameters. 

Since the true posterior $p_{\theta}(\mb{z}|\mb{A}, \mb{X}, \mb{Y}_{tr})$ is intractable, we adopt variational inference \cite{elbo} to learn it. We introduce a variational distribution $q_{\phi}(\mb{z}|\mb{A},\mb{X},\mb{Y}_{tr})$, parameterized by $\phi$, to approximate $p_{\theta}(\mb{z}|\mb{A}, \mb{X}, \mb{Y}_{tr})$, and we aim to minimize KL divergence between the two distributions: 
\begin{equation}
  \begin{split}
    \min \mathrm{KL}\left[q_{\phi}(\mb{z}|\mb{A},\mb{X},\mb{Y}_{tr}) \| p_{\theta}(\mb{z}|\mb{A}, \mb{X}, \mb{Y}_{tr}) \right]
  \end{split}
  \label{eq:kl}
\end{equation}
then following the standard derivation of variational inference (details in appendix \ref{app:der}), we can get the ELBO (Evidence Lower BOund) learning object: 
\begin{equation}
	\begin{split}
		\max \mathcal{L}_{(\theta,\phi)} = & - \mathrm{KL} \left[q_{\phi}(\mb{z}|\mb{A},\mb{X},\mb{Y}_{tr}) || p(\mb{z}) \right] + \\
    & \mathbb{E}_{q_{\phi}(\mb{z}|\mb{A},\mb{X},\mb{Y}_{tr})} \left[\log p_\theta(\mb{Y}_{tr}|\mb{z},\mb{A},\mb{X})\right] \\
    = & \,\mathcal{L}_{en} + \mathcal{L}_{de}
	\end{split}
	 \label{eq:elbo}
\end{equation}

The derived ELBO includes two terms, which respectively correspond to the encoder and decoder training of VR-GNN.  
The first term is a KL divergence, that the encoder $\phi$ is trained to generate relation vector $\mathbf{z}$ by observed graph information $\mathbf{A}$, $\mathbf{X}$, $\mathbf{Y}_{tr}$. Meanwhile, $\mathbf{z}$ is controlled by a manually assigned prior distribution $p(\mathbf{z})$. 
For the second term, the decoder $\theta$ is trained to employ generated $\mathbf{z}$, together with $\mathbf{A}$ and $\mathbf{X}$, to predict observed node labels $\mathbf{Y}_{tr}$. 
We abbreviate the two terms as $\mathcal{L}_{en}$ and $\mathcal{L}_{de}$ respectively.  

In the inference phase, the learned encoder $\phi$ can be directly used to generate relation vectors, and the decoder $\theta$ is used to predict the unknown node labels. Hence we derive the final formulization of VR-GNN:
\begin{equation}
  \begin{split}
    & p_{(\theta, \phi)}(\mathbf{Y} | \mathbf{A}, \mathbf{X}, \mb{Y}_{tr}) \\ 
    & = \int q_{\phi}(\mb{z}|\mb{A},\mb{X},\mb{Y}_{tr}) \, p_{\theta}(\mathbf{Y} | \mathbf{z}, \mathbf{A}, \mathbf{X}, \mb{Y}_{tr}) \, \mathrm{d} \mathbf{z}
  \end{split}
  \label{eq: e3}
\end{equation}
The framework demonstration can be seen in figure 2. 

\subsection{Encoder: Relation Vector Generation}
We model the relation vector of each connection as independent identically distribution, given $\mathbf{A}$, $\mathbf{X}$ and $\mathbf{Y}_{tr}$, therefore the variational posterior distribution $q_{\phi}(\mb{z}|\mb{A},\mb{X},\mb{Y}_{tr})$ and prior distribution $p(\mb{z})$ can be factorized as:
\begin{equation}
  \begin{split}
    q_{\phi}(\mb{z}|\mb{A},\mb{X},\mb{Y}_{tr}) & =\prod_{e_{ij} \in \mathcal{E}} q_{\phi}(\mb{z}_{ij}|\mb{A},\mb{X},\mb{Y}_{tr}) \\ 
    p(\mb{z}) &= \prod_{e_{ij} \in \mathcal{E}} p(\mb{z}_{ij})
  \end{split}
\end{equation}
and the learning object of encoder can be rewritten as: 
\begin{equation}
    \mathcal{L}_{en} = - \sum_{e_{ij} \in \mathcal{E}} \mathrm{KL} \left[q_{\phi}(\mb{z}_{ij}|\mb{A},\mb{X},\mb{Y}_{tr}) || p(\mb{z}_{ij})\right]
    \label{eq: encoder loss}
\end{equation}
Inspired by the implementation of VAE \cite{vae}, we let the posterior be a multivariate normal distribution, and the prior a standard multivariate normal distribution, which is flexible and could make the computation analytical: 
\begin{equation}
  \begin{split}
    q_{\phi}(\mb{z}_{ij}|\mb{A},\mb{X},\mb{Y}_{tr}) &= \mathcal{N}(\bm{\mu}_{ij}, \bm{\sigma}_{ij}^2\mb{I}) \\
    p(\mb{z}_{ij}) &= \mathcal{N}(\mb{0}, \mb{I}) \\ 
  \end{split}
\end{equation}
where $\bm{\mu}_{ij}$ and $\bm{\sigma}_{ij}^2$ are distribution parameters to be learned. 

Furthermore, the relation vector $\mb{z}_{ij}$ expects to encode the comprehensive homophily and heterophily characteristics of a connection. 
To achieve this goal, we mainly consider three aspects of information to generate $\mb{z}_{ij}$: structure, feature and label (corresponding to $\mb{A}$, $\mb{X}$ and $\mb{Y}_{tr}$), for there have been works showing graph topology, node feature and node labels to serve as important parts for homophily and heterophily modeling \cite{wrgat,DMP,cpgnn}. 


Specifically, we decompose $\mb{z}_{ij}$ into three sub-relation vectors to be separately generated from varying aspects, then linearly combined to fuse the information: 
\begin{equation}
  \mb{z}_{ij} = \alpha_s \mb{z}_{s, ij} + \alpha_f \mb{z}_{f, ij} + \alpha_l \mb{z}_{l, ij} 
  \label{eq: comb}
\end{equation}
where $\alpha\cdot$ are hyper-parameters for composing weight. $\mb{z}_{s, ij}$, $\mb{z}_{f, ij}$ and $\mb{z}_{l, ij}$ denote structure, feature and label sub-relation vectors, which like $\mb{z}$, are set as multivariate normal distribution with mutual independence:
\begin{small}
  \begin{equation}
    \begin{split}
      q_{\phi}(\mb{z}_{s, ij}|\mb{A}, \mb{X}, \mb{Y}_{tr}) & = q_{\phi}(\mb{z}_{s, ij}|\mb{A}) = \mathcal{N}(\bm{\mu}_{s, ij}, \bm{\sigma}_{s, ij}^2\mb{I}) \\
      q_{\phi}(\mb{z}_{f, ij}|\mb{A}, \mb{X}, \mb{Y}_{tr}) & = q_{\phi}(\mb{z}_{f, ij}|\mb{X}) = \mathcal{N}(\bm{\mu}_{f, ij}, \bm{\sigma}_{f, ij}^2\mb{I}) \\
      q_{\phi}(\mb{z}_{l, ij}|\mb{A}, \mb{X}, \mb{Y}_{tr}) & = q_{\phi}(\mb{z}_{l, ij}|\mb{Y}_{tr}) = \mathcal{N}(\bm{\mu}_{l, ij}, \bm{\sigma}_{l, ij}^2\mb{I})
    \end{split}
  \end{equation}
\end{small}
hence if we have got each sub-relation's expectation $\bm{\mu}_{\cdot, ij}$ and variance $\bm{\sigma}_{\cdot, ij}^2$, we can derive the distribution of $\mb{z}_{ij}$ as:
\begin{equation}
  \begin{split}
    & \bm{\mu}_{ij} = \alpha_s \bm{\mu}_{s, ij} + \alpha_f \bm{\mu}_{f, ij} + \alpha_l \bm{\mu}_{l, ij} \\
    & \bm{\sigma}_{ij}^2 = \alpha_s^2 \bm{\sigma}_{s, ij}^2 + \alpha_f^2 \bm{\sigma}_{f, ij}^2 + \alpha_l^2 \bm{\sigma}_{l, ij}^2
  \end{split}
  \label{eq: mu sigma combine}
\end{equation}
and we detail the design of each sub-relation vector below. 


\paragraph{Structure Sub-relation Vector}
For capturing structure aspect information, we assign a randomly initialized expectation and variance embedding for each connection:
\begin{equation}
  \begin{split}
    \bm{\mu}_{s, ij} \in \mathbb{R}^{|\mathcal{E}|\times H} \\
    \bm{\sigma}_{s, ij} \in \mathbb{R}^{|\mathcal{E}|\times H}
  \end{split}
\end{equation}
where $H$ denotes hidden dimension. 
Because there is no guidance for generation, the subsequent learning of sub-relation vector is in fact based on the graph structure.
This is inspired by many knowledge graph embedding works \cite{transe,pathcon}, where entity and relation embedding are normally randomly initialized, and could achieve meaningful steady state by learning triplet structure over the graph. 


\paragraph{Feature Sub-relation Vector}
This is motivated by the observation that, the feature of edge endpoints could be regarded as weak label information, and may also serve as an indicator for connection homophily and heterophily \cite{DMP}. 
Specifically, we employ the Multi-layer Perceptron (MLP) to transform the concatenation of two endpoints feature, and generate expectation and variance as follows: 
\begin{equation}
	\begin{split}
		\mb{f}_{ij} &= \mathrm{ReLU}\left(\mathrm{MLP}([\mb{x}_{i} \Vert \mb{x}_{j}])\right) \\
		\bm{\mu}_{f, ij} &= \mathrm{MLP}(\mb{f}_{ij}) \\ 
		\bm{\sigma}_{f, ij} &= \mathrm{MLP}(\mb{f}_{ij})
	\end{split}
\end{equation}
Note that the above MLPs are different modules, for reducing symbols we adopt the same denotation (the same below).

\paragraph{Label Sub-relation Vector}
The label of two end-nodes can provide direct homophily and heterophily description for a connection, but the usage of label information faces two problems: 1. There are only partially observed node labels; 2. The introduction of label information for in-degree node may lead to label leakage problem, as message passing will bring the ``correct answer'' to the node. Therefore we only use out-degree node label to generate sub-relation vector. 
Specifically, for a connection $e_{ij}$, we take node $v_i$'s label $y_i$ as one-hot vector, and feed it into MLP to generate expectation and variance embedding. For unobserved labels, we set the one-hot vector as zero. The process can be formulized as following: 
\begin{equation}
	\begin{split}
		\bm{\mu}_{l, ij} &= \mathrm{MLP}(y_i) \\  
		\bm{\sigma}_{l, ij} &= \mathrm{MLP}(y_i)
	\end{split}
\end{equation}

\paragraph{Relation Vector Generating}
After getting each sub-relations' mean and variance embedding, we combine them into the final embedding $\bm{\mu}_{ij}$ and $\bm{\sigma}_{ij}^2$ by equation \ref{eq: mu sigma combine}. 
Additionally, instead of directly sampling $\mb{z}_{ij}$, we apply the re-parameterization trick of VAE \cite{vae} to make the sampling process derivable: 
\begin{equation}
  \begin{split}
    \mb{z}_{ij} & = \bm{\mu}_{ij} + \bm{\sigma}_{ij} \epsilon
  \end{split}
\end{equation}
where $\epsilon \sim \mathcal{N}(\mb{0}, \mb{I})$. 
After getting each edge's posterior $\mb{z}_{ij}$, equation \ref{eq: encoder loss} is conducted to calculate the encoder loss. 

\subsection{Decoder: Message Passing with Relation Vector}

The decoder aims to incorporate generated relation vectors into message-passing framework, to complete downstream node classification task. 
Currently, there existing several relation-based GNN works, like R-GCN \cite{rgcn}, CompGCN \cite{CompGCN}, SE-GNN \cite{segnn}, while most works focus on knowledge graph embedding or link prediction task, few attempts have been made for graph embedding and node classification task. 

In this work, our inspiration is mainly based on the idea of TransE \cite{transe}, a classical knowledge graph model, that the relation both serves as a semantic and numerical translation for connected nodes. Our message-passing function can be formalized as follows: 
\begin{equation}
\label{eq:agg}
  \begin{gathered}
    \mb{h}_{i}^{l+1} = \mathrm{U}(\mb{h}_{i}^{l}, \mathrm{AGG}(\varphi(\mb{h}_{j}^{l}, \mb{z}_{ji}^{l}): j \in \mathcal{N}_{i})) \\ 
    \varphi(\mb{h}_{j}^{l}, \mb{z}_{ji}^{l}) = \mb{W}^{l}\mb{h}_{j}^{l}+\mb{z}_{ji}^{l} 
  \end{gathered}
\end{equation}
where $\mb{h}_{i}^{l}$ denotes the embedding of node $v_i$ in layer $l$; $\mb{z}_{ji}^{l}$ denotes the relation vector of edge $e_{ji}$ in layer $l$, with $\mb{z}_{ji}^{0}=\mb{z}_{ji}$.
For each neighbor, we apply a matrix transformation and relation translation before aggregating. This can convert neighbors to a more proper feature with the central node, and flexibly model the homophily and heterophily property of each connection when message passing. 


After each layer, relation vectors will also go through a matrix transformation, to maintain the layer consistency with node embedding:
\begin{equation}
	\begin{split}
		\mb{z}_{ji}^{l+1} &= \mb{W}_{rel}^{l} \, \mb{z}_{ji}^{l}
	\end{split}
\end{equation}

Next, we give the overall procedure of the decoder and corresponding implementation details. Firstly, we apply MLP to transform original node feature to higher-level embedding:
\begin{equation}
	\begin{split}
		\mb{h}_{i}^{0} &= \mathrm{ReLU}(\mathrm{MLP}(\mb{x}_i))
	\end{split}
	\label{eq:trans}
\end{equation}
Secondly, we conduct aggregation function. Considering that attention mechanism can adaptively model the influence of different nodes, we take self-attention \cite{attention} to aggregate neighbor information:
\begin{equation}
	\begin{split}
		\mb{\bar{h}}_{i}^{l} & = \mathrm{AGG}(\varphi(\mb{h}_{j}^{l}, \mb{z}_{ji}^{l}): j \in \mathcal{N}_{i}) \\
    & = \sum_{j\in \mathcal{N}_{i}} \beta_{ij}^{l} (\varphi(\mb{h}_{j}^{l}, \mb{z}_{ji}^{l}))
	\end{split}
\end{equation}
where $\beta_{ij}^{l}$ is attention coefficient:
\begin{equation}
	\beta_{ij}^{l} = \frac{\exp\{\mb{h}_{i}^{l}\varphi(\mb{h}_{j}^{l}, \mb{z}_{ji}^{l})\}}{\sum_{k\in \mathcal{N}_{i}}\exp\{\mb{h}_{i}^{l} \varphi(\mb{h}_{k}^{l}, \mb{z}_{ki}^{l})\}}
\end{equation}
Thirdly, we conduct updating function to renew the central node embedding:
\begin{equation}
	\begin{split}
		\mb{h}_{i}^{l+1} &= \theta \mb{\bar{h}}_{i}^{l}+(1-\theta)\mb{h}_{i}^{0} 
	\end{split}
	\label{eq:theta}
\end{equation}
where $\theta$ is to balance $\mb{\bar{h}}_{i}^{l}$ and $\mb{h}_{i}^{0}$, that can maintain the computing stability by attaching a residual of initial layer. Then the second and third steps will iterate $L$ times to get the output node representation: 
\begin{equation}
	\mb{h}_{i}=\mb{h}_{i}^{L}
\end{equation}
Finally, we employ an MLP to perform node classification: 
\begin{equation}
	\begin{split}
    y_{i}^{pred} &= \mathrm{MLP}(\mb{h}_{i})
	\end{split}
\end{equation}


\subsection{Training and Inference}

\paragraph{Training} 
After getting the prediction of each node, we can calculate a semi-supervised loss for the decoder, which corresponds to the second term of equation \ref{eq:elbo}:
\begin{equation}
	\begin{split}
		\mathcal{L}_{de} &= -\frac{1}{N_{tr}} \sum_{v_i} \mathrm{CE}(y_{i}^{pred}, \, y_{i}) \\
	\end{split}
\end{equation}
where $N_{tr}$ is the training node number; $\mathrm{CE}(\cdot)$ denotes the cross entropy function. Then with the encoder loss of equation \ref{eq: encoder loss}, we could derive the overall loss of the model: 
\begin{equation}
	\begin{split}
		\mathcal{L}_{(\theta,\phi)} =& \gamma\mathcal{L}_{en} + (1-\gamma) \mathcal{L}_{de}
	\end{split}
	\label{eq:loss}
\end{equation}
Here we add a weighting hyper-parameter $\gamma$ between the encoder and decoder, which aims to provide training process a more flexible focus. 

\paragraph{Inference}
In the inference phase, when generating $\mb{z}_{ij}$ by $q_{\phi}(\mb{z}_{ij}|\mb{A},\mb{X},\mb{Y}_{tr})$, we directly use the expectation $\bm{\mu}_{ij}$ as the relation vector of edge $e_{ij}$. We ignore the variance $\bm{\sigma}_{ij}$ to reduce the noise for inference, which is similar as \cite{gvae}.

\section{Experiments}
\subsection{Experiment Setup}

\paragraph{Datasets}
We conduct experiments on eight real-world datasets.
Among them, Cora, Citeseer and Pubmed are normally regarded as homophilic graphs, and Chameleon, Squirrel, Actor, Cornell and Texas are considered as heterophilic graphs. 
We use the same dataset partition as \cite{gprgnn}, which randomly splits nodes into train/validati on/test set with a ratio of $60\%/20\%/20\%$. 
The details of datasets can be seen in appendix \ref{app:dataset}. 

\paragraph{Baselines}
We compare VR-GNN with several state-of-the-art baselines to verify the effectiveness of our method, including: 
\textbf{MLP}: that only considers node features and ignores graph structure;
\textbf{Homophily-based GNNs}:  \textbf{GCN} \cite{gcn}, \textbf{GAT} \cite{gat} and \textbf{SGC} \cite{sgc}, which are designed with the homophily assumption;
\textbf{Heterophily-based GNNs}: \textbf{FAGCN} \cite{fagcn}, \textbf{GPR-GNN} \cite{gprgnn}, \textbf{BernNet}\cite{bernnet}, \textbf{ACM-GCN} \cite{acm}, that take the approach of passing signed messages, and \textbf{GeomGCN} \cite{geom}, \textbf{H$_2$GCN} \cite{h2gcn}, \textbf{HOC-GCN} \cite{hoc}, \textbf{BM-GCN} \cite{bmgcn}, \textbf{GloGNN++} \cite{glognn}, that take the approach of mixing high-order neighbors.
For all methods, we report the mean accuracy with a $95\%$ confidence interval of 10 runs. 
Appendix \ref{app:setting} gives more details of model settings. 

\subsection{Results of Node classification Task}
\begin{table*}[ht]
  \begin{tabular}{l|ccccc|ccc}
    \toprule[1pt]
           & \multicolumn{5}{c|}{Heterophilic Datasets}                                                                                                                                                  & \multicolumn{3}{c}{Homophilic Datasets}                                                                                                  \\
           & Chameleon                         & Squirrel                          & Actor                             & Texas                             & Cornell                           & Cora                                       & Citeseer                                & Pubmed                                  \\
  \midrule
  MLP      & {\color[HTML]{000000} 47.61$_{\pm 1.23}$} & {\color[HTML]{000000} 31.73$_{\pm 0.98}$} & {\color[HTML]{000000} 39.20$_{\pm 0.82}$} & {\color[HTML]{000000} 89.51$_{\pm 1.80}$} & {\color[HTML]{000000} 89.51$_{\pm 2.60}$}  & {\color[HTML]{000000} 77.83$_{\pm 1.28}$}          & {\color[HTML]{000000} 76.77$_{\pm 0.90}$}       & {\color[HTML]{000000} 85.67$_{\pm 0.33}$}       \\
  GCN      & {\color[HTML]{000000} 62.93$_{\pm 1.82}$} & {\color[HTML]{000000} 46.33$_{\pm 1.06}$} & {\color[HTML]{000000} 33.73$_{\pm 0.85}$} & {\color[HTML]{000000} 78.69$_{\pm 3.28}$} & {\color[HTML]{000000} 65.74$_{\pm 4.43}$} & {\color[HTML]{000000} 87.87$_{\pm 1.03}$}          & {\color[HTML]{000000} 80.26$_{\pm 0.60}$}       & {\color[HTML]{000000} 87.16$_{\pm 0.27}$}       \\
  GAT      & {\color[HTML]{000000} 63.26$_{\pm 1.40}$} & {\color[HTML]{000000} 42.81$_{\pm 1.14}$} & {\color[HTML]{000000} 35.93$_{\pm 0.42}$} & {\color[HTML]{000000} 79.67$_{\pm 2.30}$} & {\color[HTML]{000000} 77.70$_{\pm 2.62}$} & {\color[HTML]{000000} \textbf{89.14$_{\pm 0.95}$}} & {\color[HTML]{000000} 81.45$_{\pm 0.59}$}       & {\color[HTML]{000000} 87.51$_{\pm 0.25}$}       \\
  SGC      & {\color[HTML]{000000} 64.55$_{\pm 1.36}$} & {\color[HTML]{000000} 40.45$_{\pm 0.71}$} & {\color[HTML]{000000} 29.97$_{\pm 0.66}$} & {\color[HTML]{000000} 69.18$_{\pm 2.62}$} & {\color[HTML]{000000} 52.62$_{\pm 3.61}$} & {\color[HTML]{000000} 86.78$_{\pm 0.95}$}          & {\color[HTML]{000000} 80.71$_{\pm 0.55}$}       & {\color[HTML]{000000} 81.93$_{\pm 0.21}$}       \\
  
  FAGCN    & {\color[HTML]{000000} 63.30$_{\pm 1.08}$} & {\color[HTML]{000000} 41.26$_{\pm 1.24}$} & {\color[HTML]{000000} 38.36$_{\pm 0.72}$} & {\color[HTML]{000000} 90.00$_{\pm 3.78}$} & {\color[HTML]{000000} 88.38$_{\pm 2.16}$} & {\color[HTML]{000000} 87.58$_{\pm 1.09}$}          & {\color[HTML]{000000} {\ul 81.79$_{\pm 1.01}$}} & {\color[HTML]{000000} 84.26$_{\pm 0.41}$}       \\
  GPR-GNN   & {\color[HTML]{000000} 66.43$_{\pm 0.74}$} & {\color[HTML]{000000}  52.96$_{\pm 0.92}$} & {\color[HTML]{000000}  39.69$_{\pm 0.72}$} & {\color[HTML]{000000}  91.80$_{\pm 1.64}$} & {\color[HTML]{000000}  88.85$_{\pm 2.13}$} & {\color[HTML]{000000} 88.11$_{\pm 1.05}$}          & {\color[HTML]{000000} 79.51$_{\pm 0.85}$}       & {\color[HTML]{000000} 89.25$_{\pm 0.46}$}       \\
  BernNet   & {\color[HTML]{000000} 68.29$_{\pm 1.58}$} & {\color[HTML]{000000}  51.35$_{\pm 0.73}$} & {\color[HTML]{000000}  \ul 41.79$_{\pm 1.01}$} & {\color[HTML]{000000} \ul 93.12$_{\pm 0.65}$} & {\color[HTML]{000000} \ul 92.13$_{\pm 1.64}$} & {\color[HTML]{000000} 88.52$_{\pm 0.95}$}          & {\color[HTML]{000000} 80.09$_{\pm 0.79}$}       & {\color[HTML]{000000} 88.48$_{\pm 0.41}$}       \\
  ACM-GCN     & {\color[HTML]{000000} 67.74$_{\pm 1.39}$} & {\color[HTML]{000000} \ul 53.59$_{\pm 0.70}$} & {\color[HTML]{000000}  39.86$_{\pm 1.00}$} & {\color[HTML]{000000}  92.97$_{\pm 2.43}$} & {\color[HTML]{000000}  91.16$_{\pm 1,62}$} & {\color[HTML]{000000} 88.01$_{\pm 0.68}$}          & {\color[HTML]{000000} 80.87$_{\pm 0.81}$}       & {\color[HTML]{000000} 89.20$_{\pm 0.20}$}       \\
  GeomGCN   & {\color[HTML]{000000} 61.06$_{\pm 0.49}$} & {\color[HTML]{000000}  38.28$_{\pm 0.27}$} & {\color[HTML]{000000}  31.81$_{\pm 0.24}$} & {\color[HTML]{000000}  58.56$_{\pm 1.77}$} & {\color[HTML]{000000}  55.59$_{\pm 1.59}$} & {\color[HTML]{000000} 85.4$_{\pm 0.26}$}          & {\color[HTML]{000000} 76.42$_{\pm 0.37}$}       & {\color[HTML]{000000} 88.51$_{\pm 0.08}$}       \\
  H$_2$GCN   & {\color[HTML]{000000} 57.11$_{\pm 1.58}$} & {\color[HTML]{000000} 36.42 $_{\pm 1.89}$} & {\color[HTML]{000000}  35.86$_{\pm 1.03}$} & {\color[HTML]{000000}  84.86$_{\pm 6.77}$} & {\color[HTML]{000000}  82.16$_{\pm4.80}$} &  {\color[HTML]{000000} 86.92$_{\pm 1.35}$}          & {\color[HTML]{000000} 77.07$_{\pm ± 1.64}$}       & {\color[HTML]{000000} 89.40$_{\pm 0.34}$}       \\
  HOC-GCN   & - & - & {\color[HTML]{000000}  36.82$_{\pm 0.84}$} & {\color[HTML]{000000}  85.17$_{\pm 4.40}$} & {\color[HTML]{000000}  84.32$_{\pm 4.32}$} & {\color[HTML]{000000} 87.04$_{\pm 1.10}$}          & {\color[HTML]{000000} 76.15$_{\pm 1.88}$}       & {\color[HTML]{000000} 88.79$_{\pm 0.40}$}       \\
  BM-GCN   & {\color[HTML]{000000} \ul 69.85$_{\pm 0.85}$} & {\color[HTML]{000000} 51.59$_{\pm 1.05}$} & {\color[HTML]{000000} 39.23$_{\pm 0.70}$} & {\color[HTML]{000000} 83.11$_{\pm 2.79}$} & {\color[HTML]{000000} 82.79$_{\pm 2.95}$} & {\color[HTML]{000000} 87.53$_{\pm 0.70}$}          & {\color[HTML]{000000} 80.29$_{\pm 1.02}$}       & {\color[HTML]{000000} \ul 89.32$_{\pm 0.47}$}       \\
  GloGNN++ & {\color[HTML]{000000} 69.58$_{\pm 1.16}$} & {\color[HTML]{000000} 48.83$_{\pm 0.69}$} & {\color[HTML]{000000} 37.06$_{\pm 0.46}$} & {\color[HTML]{000000} 82.79$_{\pm 2.46}$} & {\color[HTML]{000000} 82.13$_{\pm 2.62}$} & {\color[HTML]{000000} 76.85$_{\pm 0.64}$}          & {\color[HTML]{000000} 75.33$_{\pm 0.78}$}       & {\color[HTML]{000000} OOM}              \\

  \midrule
  
  VR-GNN   & \textbf{71.21$_{\pm 1.17}$}               & \textbf{57.50$_{\pm 1.18}$}               & \textbf{42.16$_{\pm 0.42}$}               & \textbf{94.86$_{\pm 1.89}$}               & \textbf{92.70$_{\pm 2.70}$}               & {\ul 88.27$_{\pm 0.89}$}                           & \textbf{81.95$_{\pm 0.77}$}                     & \textbf{89.65$_{\pm 0.33}$}                     \\ \bottomrule[1pt]  
  \end{tabular}
  \caption{Results on homophilic and heterophilic datasets with mean accuracy $\boldsymbol{(\%) \pm 95\%}$ confidence interval. The best and second best results are in bold and underlined. OOM means out of memory when reproducing. H$_2$GCN and HOC-GCN report mean accuracy $\boldsymbol{(\%) \pm }$ standard deviation.}
  \label{tab:comp}
\end{table*}

\begin{figure*}[ht]
	\centering
	\subfigure[BM-GCN]{
		\includegraphics[width=0.3 \columnwidth]{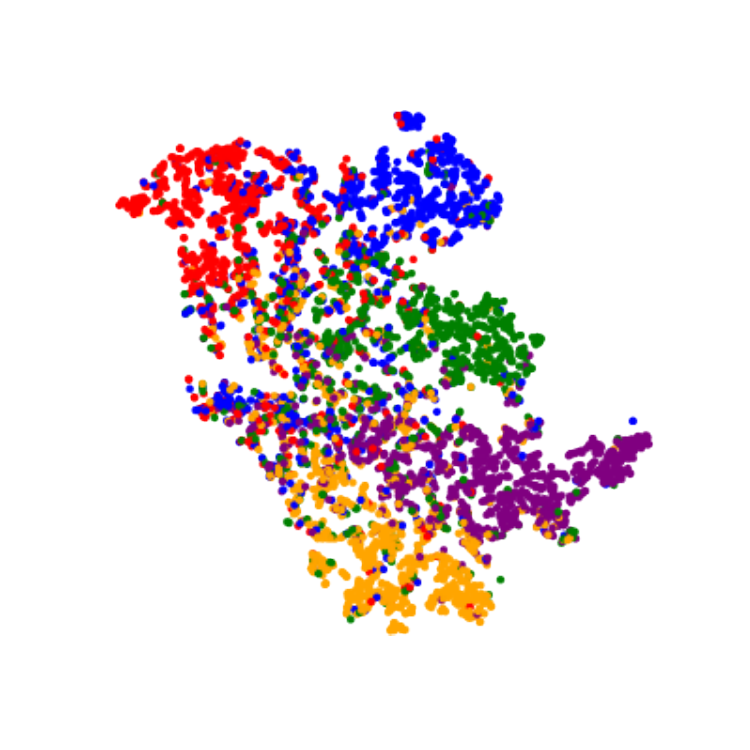}
	}
	\subfigure[GloGNN++]{
		\includegraphics[width=0.3 \columnwidth]{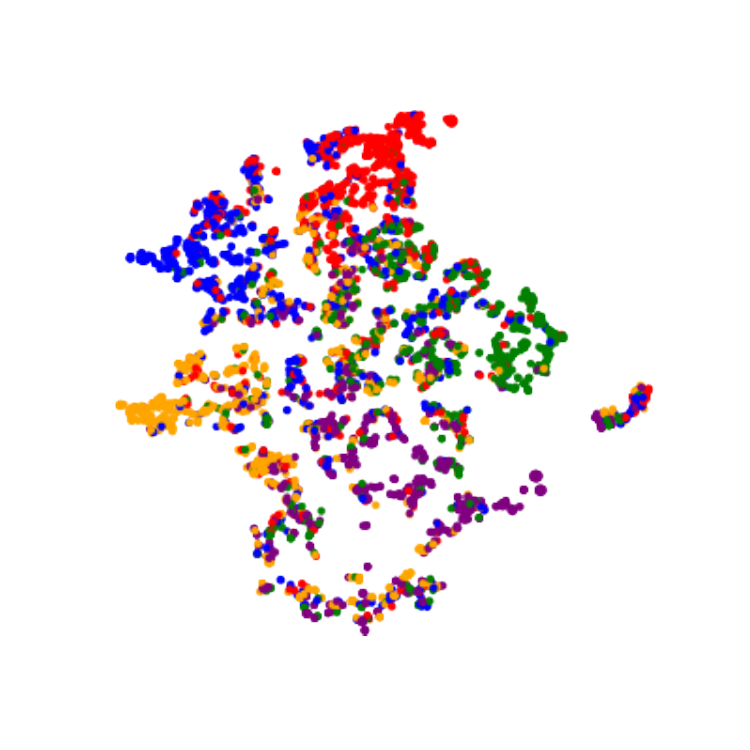}
	}
  \subfigure[FAGCN]{
		\includegraphics[width=0.3 \columnwidth]{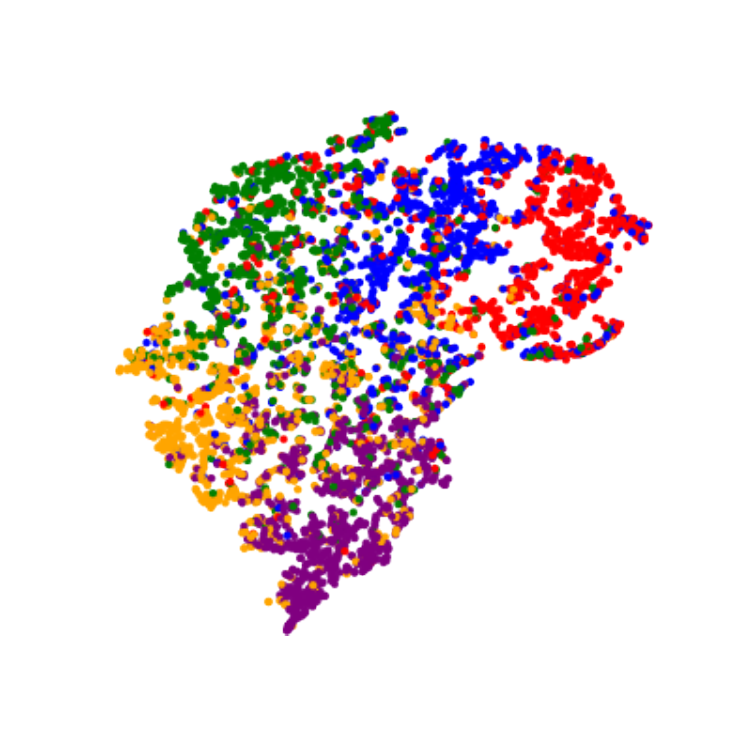}
	}
  \subfigure[GPR-GNN]{
		\includegraphics[width=0.3 \columnwidth]{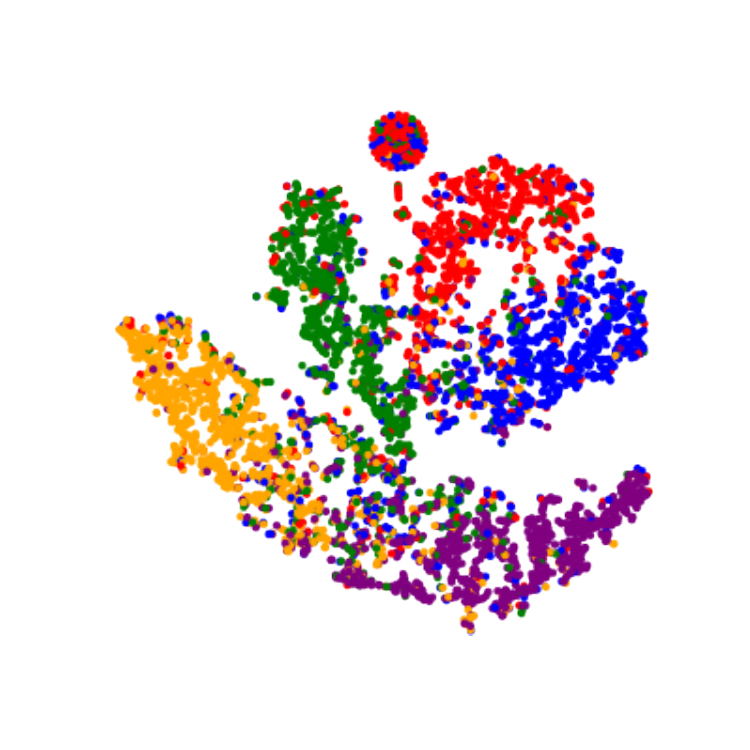}
	}
 \subfigure[ACM-GCN]{
		\includegraphics[width=0.3 \columnwidth]{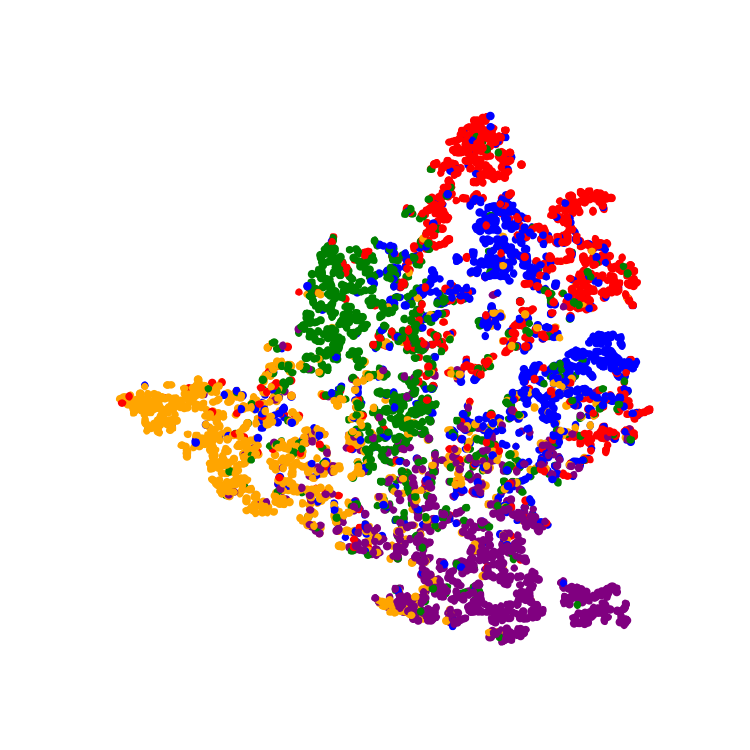}
	}
  \subfigure[VR-GNN]{
		\includegraphics[width=0.3 \columnwidth]{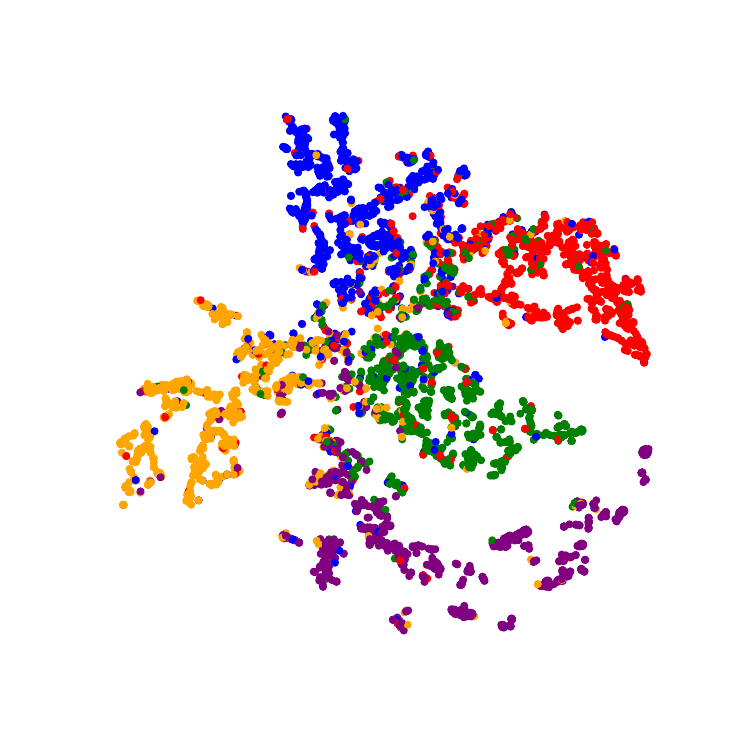}
	}
	\caption{Node embedding visualization for Squirrel dataset. Different colors correspond to different node classes.}
	\label{fig:visual}
\end{figure*}
\begin{table}[ht]
\begin{tabular}{l|cccc}
\toprule[1pt]
Datasets   & Chameleon & Squirrel & Texas & Citeseer \\
\midrule
VR-GNN$_s$  & 69.89     & 53.37    & 93.24 & 81.02    \\
VR-GNN$_f$  & 69.54     & 52.74    & 93.51 & 81.16    \\
VR-GNN$_l$  & 70.00     & 53.43    & 92.97 & 80.96    \\
VR-GNN$_sf$ & 70.22     & 55.12    & 93.71 & 81.36    \\
VR-GNN$_sl$ & 70.69     & 56.78    & 93.71 & 81.20    \\
VR-GNN$_fl$ & 70.30     & 54.48    & 93.51 & 81.88    \\
\midrule
VR-GNN     & \textbf{71.21}     & \textbf{57.50}    & \textbf{94.86} & \textbf{81.95}    \\ \bottomrule[1pt]
\end{tabular}
\caption{Ablation study of three sub-relations.}
\label{tab:ablation}
\end{table}


Table \ref{tab:comp} lists the results of VR-GNN and other baselines for node classification task, from which we can observe that:

VR-GNN outperforms all the other methods on all five heterophilic datasets. 
This proves the effectiveness of employing relation vectors to achieve a heterophily-based GNN.
Specifically, VR-GNN significantly outperforms traditional GNNs, i.e. GCN, GAT and SGC, relatively by $24.8\%$, $20.5\%$ and $41.3\%$ on average, since they cannot generalize to heterophily scenarios.
Compared with other heterophily-based GNNs, including both the method type of mixing high-order neighbors and passing signed messages, VR-GNN also achieves effective improvements, like $5.1\%$ over ACM-GNN on Chameleon, $11.8\%$ over BernNet on Squirrel, $13.8\%$ over GloGNN++ on Actor, $14.1\%$ over BM-GCN on Texas, $4.3\%$ over GPR-GNN on Cornell. 
These results demonstrate that VR-GNN could model the heterophilic connections of the graph more flexibly and expressively, meanwhile without destroying the graph structure.

On homophilic datasets, i.e. Cora, Citeseer and Pubmed, VR-GNN performs better or comparably to the baselines. Specifically, VR-GNN outperforms all the methods on Citeseer and Pubmed dataset. For Cora dataset, VR-GNN also achieves the second best report with only $0.87$ difference with GAT. These show that VR-GNN possesses a consistent performance on homophily scenarios, which further proves the adaptive modeling capacity of relation vectors.


\subsection{Ablation Study of Three Sub-relations}
To evaluate the effect of each sub-relation vector part, we do the ablation study of only removing one sub-relation part and simultaneously removing two of them. We take Chameleon, Squirrel, Texas and Citeseer as example datasets. The results are demonstrated in table \ref{tab:ablation}. The subscripts $s$, $f$, $l$ respectively denotes the sub-relation used.
 We can see that generating relation vectors with absence of some relation cannot provide stable performance across datasets compared to VR-GNN, which verifies the necessity of composing all three parts.


\subsection{Visualization Analysis}
To show the modeling effect of VR-GNN more intuitively, we conduct the node embedding visualization for Squirrel dataset. 
We extract the node embedding of VR-GNN and five state-of-the-art baselines (BM-GCN, GloGNN++, FAGCN, GPR-GNN and ACM-GCN), then employ t-SNE \cite{tsne} algorithm to map them into 2-dimensional space for visualization. The results are shown in figure \ref{fig:visual}.
We can observe that VR-GNN achieves more discriminative node embedding, which is more cohesive within the same category and dispersed between the different categories. 
This further proves the validity of the relation vector based message-passing, which can produce more accurate assimilation and dissimilation effect between nodes according to homophily and heterophily connections.

Additionally, we also conduct the visualization analysis for relation vectors, which is placed in appendix \ref{app:rel} due to space limitation.

\subsection{Hyper-parameter Analysis}
In this section, we investigate the sensitivity of hyper-parameters used in VR-GNN. We take Chameleon, Squirrel, Cora and Pubmed as example datasets.

\begin{figure}[h]
	\centering
    \captionsetup[subfigure]{labelformat=empty}
	\subfigure{
		\includegraphics[width=0.47 \columnwidth]{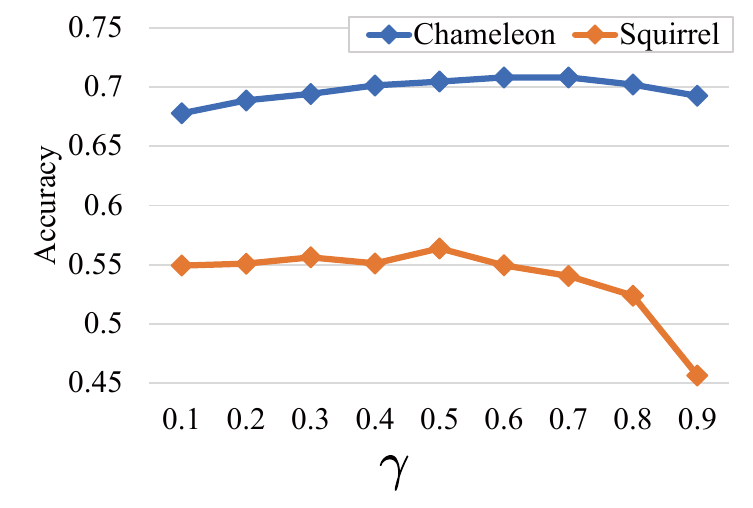}
	}
	\subfigure{
		\includegraphics[width=0.47 \columnwidth]{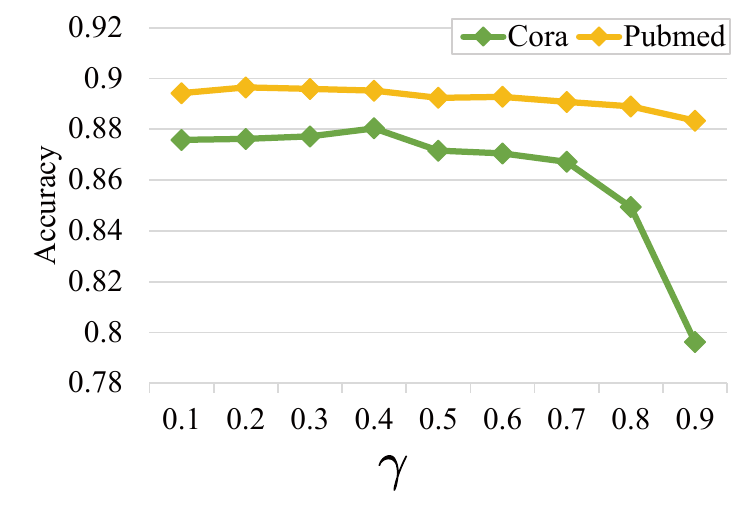}
	}
	\caption{Influence of weight parameter $\gamma$ on Chameleon, Squirrel, Cora and Pubmed dataset. }
	\label{fig:gamma}
\end{figure}

\begin{figure}[h]
	\centering
  \captionsetup[subfigure]{labelformat=empty}
	\subfigure{
		\includegraphics[width=0.47 \columnwidth]{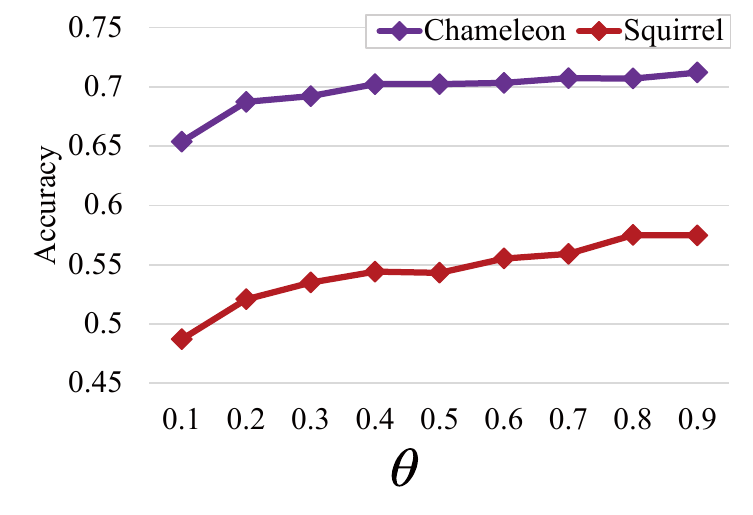}
	}
	\subfigure{
		\includegraphics[width=0.47 \columnwidth]{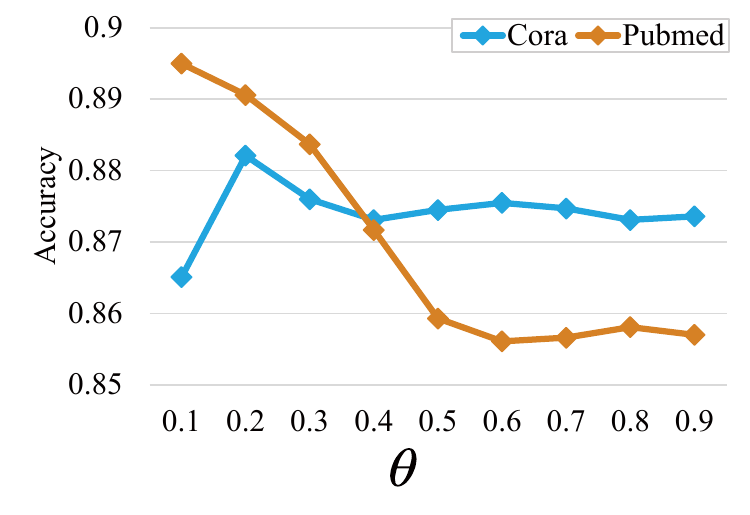}
	}
	\caption{Influence of weight parameter $\theta$ on Chameleon, Squirrel, Cora and Pubmed dataset. 
  }
	\label{fig:theta}
\end{figure}

\begin{figure}[t]
	\centering
		\includegraphics[width=0.8 \columnwidth]{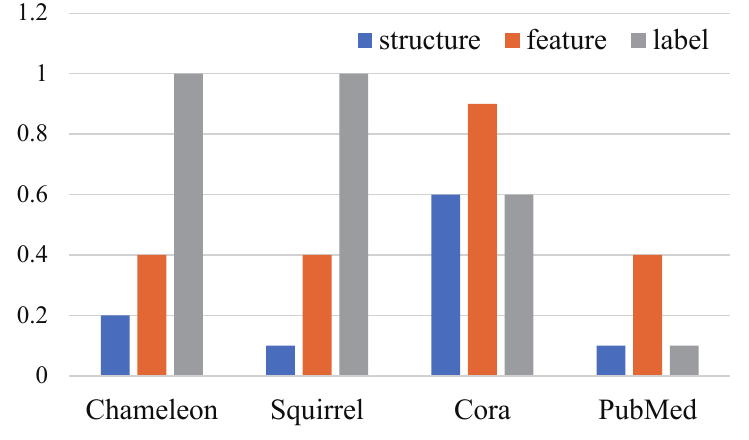}
	\caption{Analysis of parameters $\alpha_s,\alpha_f,\alpha_l$. }
	\label{fig:alpha}
\end{figure}

\paragraph{Weight Parameter $\gamma$.}
To investigate the influence of KL divergence loss (encoder loss) for learning effect, we conduct the sensitivity experiment for parameter $\gamma$ (equation \ref{eq:loss}). 
We test the node classification accuracy of VR-GNN with $\gamma$ ranging from $0.1$ to $0.9$. The results are reported in figure \ref{fig:gamma}.
We can discover that the trends of $\gamma$ are the same in all datasets, which from a low point slowly rise to a maximum and then gradually decline. This is because KL loss restricts the generated relation vector not to deviating far from the prior distribution, and the small $\gamma$ may lead to too large or too small embedding, while the large $\gamma$ will harm the learning of classification task. 
\paragraph{Weight Parameter $\theta$.}
The parameter $\theta$ balances the node's original feature and neighbor aggregation information (equation \ref{eq:theta}).  
A larger $\theta$ indicates a greater role graph structure plays. 
We test the node classification accuracy of VR-GNN with $\theta$ from $0.1$ to $0.9$. The results are shown in figure \ref{fig:theta}. 
We can observe that VR-GNN has greater $\theta$ on Chameleon and Squirrel datasets. 
This is because in Chameleon and Squirrel graph structure is more important, while in Cora and Pubmed node feature is more important. This can also be proven in table \ref{tab:comp}: VR-GNN improves $49.6\%$ and $81.2\%$ over MLP on Chameleon and Squirrel, while only $13.4\%$ and $4.6\%$ on Cora and Pubmed.
\paragraph{Analysis of Parameters $\alpha_s,\alpha_f,\alpha_l$.}
Figure \ref{fig:alpha} shows the best results of $\alpha_s$, $\alpha_f$ and $\alpha_l$ on four datasets. We can observe that although the parameters of different datasets are not completely consistent, they still show some similarity, which is related to the characteristics of corresponding datasets. Previous work \cite{ana_hete} has shown that the label distribution of Chameleon and Squirrel greatly improves the results, so the weights of $\alpha_l$ are the largest on these two datasets. Cora and PubMed have better initial feature, so $\alpha_f$ has more weight.This further illustrates the necessity of combining all three relations to achieve stable performance across datasets.
\label{app:an}

\section{Related Work}
According to the connection characteristics of the applied graphs, we introduce two families of GNN works: homophily-based GNNs and heterophily-based GNNs.

The homophily-based GNNs are developed mainly for homophilic graphs, which are the earliest proposed and extensively studied. 
GCN \cite{gcn} proposes an expressive and efficient graph convolution paradigm by simplifying the polynomial convolution kernel; 
GAT \cite{gat} applies the self-attention mechanism to adaptively adjust aggregation weights;
SGC \cite{sgc} designs a lightweight GNN by disentangling convolutional filters and weight matrices;
GraphSage \cite{sage} proposes  an inductive node embedding method with sampling and aggregating features from neighborhood.
GIN \cite{gin} designs a simple yet effective convolution mechanism to explore the upper bound of message-passing based GNNs under homophily.

The poor performance of homophily-based GNNs on heterophilic scenarios inspires the study of heterophily-based GNNs. 
Geom-GCN \cite{geom} proposes a novel geometric aggregation scheme to acquire more homophilic neighbors; 
BM-GCN \cite{bmgcn} explores block-guided neighbors and conducts classified aggregation for both homophilic and heterophilic nodes; 
GloGNN \cite{glognn} learns a coefficient matrix from graph and utilizes it to aggregate nodes with global homophily;
FAGCN \cite{fagcn} proposes an adaptive method to capture both low and high-frequency graph signal by passing signed message; 
GPR-GNN \cite{gprgnn} learns signed weighting of different orders of graph structure to deal with both homophily and heterophily.

\section{Conclusion}
In this paper, we propose a new method that utilizes relation vectors to model homophily and heterophily. Then we propose a novel model named \textbf{V}ariational \textbf{R}elation Vector \textbf{G}raph \textbf{N}eural \textbf{N}etwork (\textbf{VR-GNN}). 
VR-GNN builds the framework based on Variational Auto-Encoder (VAE) and provides an effective end-to-end solution for both relation vector generation and relation fused message-passing.
Extensive experiments on eight real-world datasets verify the validity of our method.

\bibliographystyle{named}
\bibliography{reference}

\begin{thebibliography}{}

\bibitem[\protect\citeauthoryear{Bahdanau \bgroup \em et al.\egroup
  }{2015}]{attention}
Dzmitry Bahdanau, {Kyung Hyun} Cho, and Yoshua Bengio.
\newblock Neural machine translation by jointly learning to align and
  translate.
\newblock January 2015.
\newblock 3rd International Conference on Learning Representations, ICLR 2015 ;
  Conference date: 07-05-2015 Through 09-05-2015.

\bibitem[\protect\citeauthoryear{Bo \bgroup \em et al.\egroup }{2021}]{fagcn}
Deyu Bo, Xiao Wang, Chuan Shi, and Huawei Shen.
\newblock Beyond low-frequency information in graph convolutional networks.
\newblock In {\em Proceedings of the AAAI Conference on Artificial
  Intelligence}, volume~35, pages 3950--3957, 2021.

\bibitem[\protect\citeauthoryear{Bordes \bgroup \em et al.\egroup
  }{2013}]{transe}
Antoine Bordes, Nicolas Usunier, Alberto Garcia-Duran, Jason Weston, and Oksana
  Yakhnenko.
\newblock Translating embeddings for modeling multi-relational data.
\newblock {\em Advances in neural information processing systems}, 26, 2013.

\bibitem[\protect\citeauthoryear{Chen \bgroup \em et al.\egroup }{2020}]{gcnii}
Ming Chen, Zhewei Wei, Zengfeng Huang, Bolin Ding, and Yaliang Li.
\newblock Simple and deep graph convolutional networks.
\newblock In {\em International Conference on Machine Learning}, pages
  1725--1735. PMLR, 2020.

\bibitem[\protect\citeauthoryear{Chien \bgroup \em et al.\egroup
  }{2020}]{gprgnn}
Eli Chien, Jianhao Peng, Pan Li, and Olgica Milenkovic.
\newblock Adaptive universal generalized pagerank graph neural network.
\newblock In {\em International Conference on Learning Representations}, 2020.

\bibitem[\protect\citeauthoryear{Ekambaram}{2014}]{hete_book}
Venkatesan~Nallampatti Ekambaram.
\newblock {\em Graph-structured data viewed through a Fourier lens}.
\newblock University of California, Berkeley, 2014.

\bibitem[\protect\citeauthoryear{Fey and Lenssen}{2019}]{pyg}
Matthias Fey and Jan~E. Lenssen.
\newblock Fast graph representation learning with {PyTorch Geometric}.
\newblock In {\em ICLR Workshop on Representation Learning on Graphs and
  Manifolds}, 2019.

\bibitem[\protect\citeauthoryear{Gilmer \bgroup \em et al.\egroup
  }{2017}]{messagepassing}
Justin Gilmer, Samuel~S Schoenholz, Patrick~F Riley, Oriol Vinyals, and
  George~E Dahl.
\newblock Neural message passing for quantum chemistry.
\newblock In {\em International conference on machine learning}, pages
  1263--1272. PMLR, 2017.

\bibitem[\protect\citeauthoryear{Hamilton \bgroup \em et al.\egroup
  }{2017}]{sage}
Will Hamilton, Zhitao Ying, and Jure Leskovec.
\newblock Inductive representation learning on large graphs.
\newblock {\em Advances in neural information processing systems}, 30, 2017.

\bibitem[\protect\citeauthoryear{He \bgroup \em et al.\egroup }{2021}]{bernnet}
Mingguo He, Zhewei Wei, Zengfeng Huang, and Hongteng Xu.
\newblock Bernnet: Learning arbitrary graph spectral filters via bernstein
  approximation.
\newblock In {\em NeurIPS}, 2021.

\bibitem[\protect\citeauthoryear{He \bgroup \em et al.\egroup }{2022}]{bmgcn}
Dongxiao He, Chundong Liang, Huixin Liu, Mingxiang Wen, Pengfei Jiao, and
  Zhiyong Feng.
\newblock Block modeling-guided graph convolutional neural networks.
\newblock In {\em Proceedings of the AAAI Conference on Artificial
  Intelligence}, volume~36, pages 4022--4029, 2022.

\bibitem[\protect\citeauthoryear{Hinton and van~der Maaten}{2008}]{tsne}
G~Hinton and LJP van~der Maaten.
\newblock Visualizing data using t-sne journal of machine learning research.
\newblock 2008.

\bibitem[\protect\citeauthoryear{Hogan \bgroup \em et al.\egroup }{2021}]{kg}
Aidan Hogan, Eva Blomqvist, Michael Cochez, Claudia d'Amato, Gerard~de Melo,
  Claudio Gutierrez, Sabrina Kirrane, Jos{\'e} Emilio~Labra Gayo, Roberto
  Navigli, Sebastian Neumaier, et~al.
\newblock Knowledge graphs.
\newblock {\em Synthesis Lectures on Data, Semantics, and Knowledge},
  12(2):1--257, 2021.

\bibitem[\protect\citeauthoryear{Kingma and Welling}{2014}]{vae}
Diederik~P. Kingma and Max Welling.
\newblock {Auto-Encoding Variational Bayes}.
\newblock In {\em 2nd International Conference on Learning Representations,
  {ICLR} 2014, Banff, AB, Canada, April 14-16, 2014, Conference Track
  Proceedings}, 2014.

\bibitem[\protect\citeauthoryear{Kipf and Welling}{2016}]{gvae}
Thomas~N Kipf and Max Welling.
\newblock Variational graph auto-encoders.
\newblock {\em NIPS Workshop on Bayesian Deep Learning}, 2016.

\bibitem[\protect\citeauthoryear{Kipf and Welling}{2017}]{gcn}
Thomas~N. Kipf and Max Welling.
\newblock Semi-supervised classification with graph convolutional networks.
\newblock In {\em International Conference on Learning Representations (ICLR)},
  2017.

\bibitem[\protect\citeauthoryear{Kipf \bgroup \em et al.\egroup }{2018}]{RI}
Thomas Kipf, Ethan Fetaya, Kuan-Chieh Wang, Max Welling, and Richard Zemel.
\newblock Neural relational inference for interacting systems.
\newblock In {\em International Conference on Machine Learning}, pages
  2688--2697. PMLR, 2018.

\bibitem[\protect\citeauthoryear{Klicpera \bgroup \em et al.\egroup
  }{2018}]{appnp}
Johannes Klicpera, Aleksandar Bojchevski, and Stephan G{\"u}nnemann.
\newblock Predict then propagate: Graph neural networks meet personalized
  pagerank.
\newblock In {\em International Conference on Learning Representations}, 2018.

\bibitem[\protect\citeauthoryear{Li \bgroup \em et al.\egroup }{2018}]{deeper}
Qimai Li, Zhichao Han, and Xiao-Ming Wu.
\newblock Deeper insights into graph convolutional networks for semi-supervised
  learning.
\newblock In {\em Thirty-Second AAAI conference on artificial intelligence},
  2018.

\bibitem[\protect\citeauthoryear{Li \bgroup \em et al.\egroup }{2022a}]{segnn}
Ren Li, Yanan Cao, Qiannan Zhu, Guanqun Bi, Fang Fang, Yi~Liu, and Qian Li.
\newblock How does knowledge graph embedding extrapolate to unseen data: a
  semantic evidence view.
\newblock In {\em Proceedings of the AAAI Conference on Artificial
  Intelligence}, volume~36, pages 5781--5791, 2022.

\bibitem[\protect\citeauthoryear{Li \bgroup \em et al.\egroup }{2022b}]{glognn}
Xiang Li, Renyu Zhu, Yao Cheng, Caihua Shan, Siqiang Luo, Dongsheng Li, and
  Weining Qian.
\newblock Finding global homophily in graph neural networks when meeting
  heterophily.
\newblock {\em arXiv preprint arXiv:2205.07308}, 2022.

\bibitem[\protect\citeauthoryear{Lim \bgroup \em et al.\egroup }{2021}]{hete1}
Derek Lim, Felix Hohne, Xiuyu Li, Sijia~Linda Huang, Vaishnavi Gupta, Omkar
  Bhalerao, and Ser~Nam Lim.
\newblock Large scale learning on non-homophilous graphs: New benchmarks and
  strong simple methods.
\newblock {\em Advances in Neural Information Processing Systems}, 34, 2021.

\bibitem[\protect\citeauthoryear{Luan \bgroup \em et al.\egroup }{2022}]{acm}
Sitao Luan, Chenqing Hua, Qincheng Lu, Jiaqi Zhu, Mingde Zhao, Shuyuan Zhang,
  Xiao-Wen Chang, and Doina Precup.
\newblock Revisiting heterophily for graph neural networks.
\newblock {\em Conference on Neural Information Processing Systems}, 2022.

\bibitem[\protect\citeauthoryear{Ma \bgroup \em et al.\egroup
  }{2022}]{ana_hete}
Yao Ma, Xiaorui Liu, Neil Shah, and Jiliang Tang.
\newblock Is homophily a necessity for graph neural networks?
\newblock In {\em The Tenth International Conference on Learning
  Representations, {ICLR} 2022, Virtual Event, April 25-29, 2022}.
  OpenReview.net, 2022.

\bibitem[\protect\citeauthoryear{McPherson \bgroup \em et al.\egroup
  }{2001}]{homobook}
Miller McPherson, Lynn Smith-Lovin, and James~M Cook.
\newblock Birds of a feather: Homophily in social networks.
\newblock {\em Annual Review of Sociology}, 27(1):415--444, 2001.

\bibitem[\protect\citeauthoryear{Mnih and Gregor}{2014}]{elbo}
Andriy Mnih and Karol Gregor.
\newblock Neural variational inference and learning in belief networks.
\newblock In {\em International Conference on Machine Learning}, pages
  1791--1799. PMLR, 2014.

\bibitem[\protect\citeauthoryear{Namata \bgroup \em et al.\egroup
  }{2012}]{citation_dataset2}
Galileo Namata, Ben London, Lise Getoor, Bert Huang, and UMD EDU.
\newblock Query-driven active surveying for collective classification.
\newblock In {\em 10th International Workshop on Mining and Learning with
  Graphs}, volume~8, page~1, 2012.

\bibitem[\protect\citeauthoryear{Pei \bgroup \em et al.\egroup }{2019}]{geom}
Hongbin Pei, Bingzhe Wei, Kevin Chen-Chuan Chang, Yu~Lei, and Bo~Yang.
\newblock Geom-gcn: Geometric graph convolutional networks.
\newblock In {\em International Conference on Learning Representations}, 2019.

\bibitem[\protect\citeauthoryear{Sanyal \bgroup \em et al.\egroup
  }{2020}]{protein}
Soumya Sanyal, Ivan Anishchenko, Anirudh Dagar, David Baker, and Partha
  Talukdar.
\newblock Proteingcn: Protein model quality assessment using graph
  convolutional networks.
\newblock {\em bioRxiv}, 2020.

\bibitem[\protect\citeauthoryear{Schlichtkrull \bgroup \em et al.\egroup
  }{2018}]{rgcn}
Michael Schlichtkrull, Thomas~N Kipf, Peter Bloem, Rianne van~den Berg, Ivan
  Titov, and Max Welling.
\newblock Modeling relational data with graph convolutional networks.
\newblock In {\em European semantic web conference}, pages 593--607. Springer,
  2018.

\bibitem[\protect\citeauthoryear{Sen \bgroup \em et al.\egroup
  }{2008}]{citation_dataset1}
Prithviraj Sen, Galileo Namata, Mustafa Bilgic, Lise Getoor, Brian Galligher,
  and Tina Eliassi-Rad.
\newblock Collective classification in network data.
\newblock {\em AI magazine}, 29(3):93--93, 2008.

\bibitem[\protect\citeauthoryear{Suresh \bgroup \em et al.\egroup
  }{2021}]{wrgat}
Susheel Suresh, Vinith Budde, Jennifer Neville, Pan Li, and Jianzhu Ma.
\newblock Breaking the limit of graph neural networks by improving the
  assortativity of graphs with local mixing patterns.
\newblock In {\em KDD}, pages 1541--1551, 2021.

\bibitem[\protect\citeauthoryear{Vashishth \bgroup \em et al.\egroup
  }{2019}]{CompGCN}
Shikhar Vashishth, Soumya Sanyal, Vikram Nitin, and Partha Talukdar.
\newblock Composition-based multi-relational graph convolutional networks.
\newblock In {\em International Conference on Learning Representations}, 2019.

\bibitem[\protect\citeauthoryear{Veli{\v{c}}kovi{\'{c}} \bgroup \em et
  al.\egroup }{2018}]{gat}
Petar Veli{\v{c}}kovi{\'{c}}, Guillem Cucurull, Arantxa Casanova, Adriana
  Romero, Pietro Li{\`{o}}, and Yoshua Bengio.
\newblock {Graph Attention Networks}.
\newblock {\em International Conference on Learning Representations}, 2018.
\newblock accepted as poster.

\bibitem[\protect\citeauthoryear{Wang \bgroup \em et al.\egroup
  }{2019}]{social2}
Hao Wang, Tong Xu, Qi~Liu, Defu Lian, Enhong Chen, Dongfang Du, Han Wu, and Wen
  Su.
\newblock Mcne: an end-to-end framework for learning multiple conditional
  network representations of social network.
\newblock In {\em KDD}, pages 1064--1072, 2019.

\bibitem[\protect\citeauthoryear{Wang \bgroup \em et al.\egroup
  }{2021}]{pathcon}
Hongwei Wang, Hongyu Ren, and Jure Leskovec.
\newblock Relational message passing for knowledge graph completion.
\newblock In {\em KDD}, pages 1697--1707, 2021.

\bibitem[\protect\citeauthoryear{Wang \bgroup \em et al.\egroup }{2022}]{hoc}
Tao Wang, Di~Jin, Rui Wang, Dongxiao He, and Yuxiao Huang.
\newblock Powerful graph convolutional networks with adaptive propagation
  mechanism for homophily and heterophily.
\newblock In {\em Proceedings of the AAAI Conference on Artificial
  Intelligence}, volume~36, pages 4210--4218, 2022.

\bibitem[\protect\citeauthoryear{Wu \bgroup \em et al.\egroup }{2019}]{sgc}
Felix Wu, Amauri Souza, Tianyi Zhang, Christopher Fifty, Tao Yu, and Kilian
  Weinberger.
\newblock Simplifying graph convolutional networks.
\newblock In {\em International conference on machine learning}, pages
  6861--6871. PMLR, 2019.

\bibitem[\protect\citeauthoryear{Xu \bgroup \em et al.\egroup }{2018}]{gin}
Keyulu Xu, Weihua Hu, Jure Leskovec, and Stefanie Jegelka.
\newblock How powerful are graph neural networks?
\newblock In {\em International Conference on Learning Representations}, 2018.

\bibitem[\protect\citeauthoryear{Yang \bgroup \em et al.\egroup }{2021}]{DMP}
Liang Yang, Mengzhe Li, Liyang Liu, Chuan Wang, Xiaochun Cao, Yuanfang Guo,
  et~al.
\newblock Diverse message passing for attribute with heterophily.
\newblock {\em Advances in Neural Information Processing Systems}, 34, 2021.

\bibitem[\protect\citeauthoryear{Zhu \bgroup \em et al.\egroup }{2020}]{h2gcn}
Jiong Zhu, Yujun Yan, Lingxiao Zhao, Mark Heimann, Leman Akoglu, and Danai
  Koutra.
\newblock Beyond homophily in graph neural networks: Current limitations and
  effective designs.
\newblock {\em Advances in Neural Information Processing Systems},
  33:7793--7804, 2020.

\bibitem[\protect\citeauthoryear{Zhu \bgroup \em et al.\egroup }{2021}]{cpgnn}
Jiong Zhu, Ryan~A Rossi, Anup Rao, Tung Mai, Nedim Lipka, Nesreen~K Ahmed, and
  Danai Koutra.
\newblock Graph neural networks with heterophily.
\newblock In {\em Proceedings of the AAAI Conference on Artificial
  Intelligence}, volume~35, pages 11168--11176, 2021.

\end{thebibliography}

\newpage

\begin{table*}[ht]
	\begin{tabular}{lcccccccc}
		\toprule[1pt]
		& Cora  & CiteSeer & PubMed & Chameleon & Squirrel & Actor & Texas & Cornell \\
		\midrule
		\qquad \#\,\:Node           & 2708  & 3327     & 19717  & 2277      & 5201     & 7600  & 183   & 183     \\
		\qquad \#\,\:Edge           & 10556 & 9228     & 88651  & 62742     & 396706   & 53318 & 558   & 554     \\
		\qquad \#\,\:Feature & 1433  & 3703     & 500    & 767       & 2089     & 932   & 1703  & 1703    \\
		\qquad \#\,\:Class          & 7     & 6        & 5      & 5         & 5        & 5     & 5     & 5  \\
		Homophily Ratio $\mathcal{H}$ \quad         & 0.656     & 0.578        & 0.644      & 0.024         & 0.055        & 0.008     & 0.016     & 0.137    \\ \bottomrule[1pt]  
  \end{tabular}
  \centering
  \caption{Dataset Statistics}
  \label{tab:dataset}
\end{table*}
\newpage
\appendix

\section{ELBO Derivation}
\label{app:der}
We start from the KL divergence of equation \ref{eq:kl}, which can be transformed with following steps: 
\begin{equation*}
    \begin{split}
        & \, \mathrm{KL}\left[q_{\phi}(\mb{z}|\mb{A},\mb{X},\mb{Y}_{tr}) \| p_{\theta}(\mb{z}|\mb{A}, \mb{X}, \mb{Y}_{tr}) \right] \\
        = & \, \mathbf{E}_{\mb{z}\sim q_{\phi}}\left[\log q_{\phi}(\mb{z}|\mb{A},\mb{X},\mb{Y}_{tr}) - \log p_{\theta}(\mb{z}|\mb{A}, \mb{X}, \mb{Y}_{tr})\right] \\
        = & \, \mathbf{E}_{\mb{z}\sim q_{\phi}}\left[\log q_{\phi}(\mb{z}|\mb{A},\mb{X},\mb{Y}_{tr})- \log \frac{p_{\theta}(\mb{z}, \mb{Y}_{tr}|\mb{A}, \mb{X})}{p_\theta(\mb{Y}_{tr}|\mb{A},\mb{X})}\right] \\
        = & \, \mathbf{E}_{\mb{z}\sim q_{\phi}}\left[\log q_{\phi}(\mb{z}|\mb{A},\mb{X},\mb{Y}_{tr})-\log p_{\theta}(\mb{z}, \mb{Y}_{tr}|\mb{A}, \mb{X})\right] + \\
        & \, \mathbf{E}_{\mb{z}\sim q_{\phi}}[p_\theta(\mb{Y}_{tr}|\mb{A},\mb{X})] \\
        = & -\mathcal{L}_{(\theta,\phi)}+p_\theta(\mb{Y}_{tr}|\mb{A},\mb{X}) \geq 0
    \end{split}
\end{equation*}
Therefore, maximizing the log-likelihood $p_\theta(\mb{Y}_{tr}|\mb{A},\mb{X})$ is equivalent to maximizing its lower bound, i.e. ELBO:
\begin{equation*}
    \begin{split}
        \mathcal{L}_{(\theta,\phi)} = 
        & -\mathbf{E}_{\mb{z}\sim q_{\phi}}\left[\log q_{\phi}(\mb{z}|\mb{A},\mb{X},\mb{Y}_{tr}) - \log p_{\theta}(\mb{z}, \mb{Y}_{tr}|\mb{A}, \mb{X})\right] \\
        = & \, \mathbf{E}_{\mb{z}\sim q_{\phi}}\left[-(\log q_{\phi}(\mb{z}|\mb{A},\mb{X},\mb{Y}_{tr})-\log p(\mb{z}))\right] + \\
        & \, \mathbf{E}_{\mb{z}\sim q_{\phi}}\left[\log p_{\theta}(\mb{z}, \mb{Y}_{tr}|\mb{A}, \mb{X})-\log p(\mb{z})\right] \\
        = & - \mathrm{KL} \left[q_{\phi}(\mb{z}|\mb{A},\mb{X},\mb{Y}_{tr}) || p(\mb{z}) \right] + \\ 
        & \mathbb{E}_{q_{\phi}(\mb{z}|\mb{A},\mb{X},\mb{Y}_{tr})} \left[\log p_\theta(\mb{Y}_{tr}|\mb{z},\mb{A},\mb{X})\right] \\
        = & \, \mathcal{L}_{en} + \mathcal{L}_{de}
    \end{split}
\end{equation*}

\section{Algorithm Complexity Analysis}
\label{app:com}
Here we analyse the time complexity of VR-GNN. 

The generation of each sub-relation vector costs $\mathcal{O}(|\mathcal{E}|)$, hence the time complexity of encoder is $\mathcal{O}(|\mathcal{E}|)$. 
For the decoder, the feature transformation in equation \ref{eq:trans} is of $\mathcal{O}(|\mathcal{V}|)$ complexity. For each GNN layer, the aggregation function and updating function respectively cost $\mathcal{O}(|\mathcal{E}|)$ and $\mathcal{O}(|\mathcal{V}|)$ complexity. 
Finally, the MLP for classification is of $\mathcal{O}(|\mathcal{V}|)$ complexity.

Therefore, the overall time complexity of VR-GNN is:
\begin{equation*}
    \begin{split}
        \mathcal{O}(L|\mathcal{E}|+L|\mathcal{V}|)
    \end{split}
\end{equation*}
where $L$ is the layer number. This matches the complexity degree of other GNN baselines, like FAGCN \cite{fagcn} and GPR-GNN \cite{gprgnn}.

\section{Dataset Details}
\label{app:dataset}
In the experiments, we utilize eight real-world datasets with different homophily ratio: 
\begin{itemize}
  \item Cora, Citeseer and Pubmed are three citation networks \cite{citation_dataset1,citation_dataset2} with high homophily ratio.
  \item Chameleon and Squirrel \cite{geom} are page-page networks extracted from Wikipedia of specific topics, with low homophily ratio.
  \item Actor \cite{geom} is constructed according to the actor co-occurrence in Wikipedia pages and holds low homophily ratio. 
  \item Cornell and Texas \cite{geom} are two sub-datasets of WebKB, a webpage database constructed by  Carnegie Mellon University, and possess low homophily ratio.
\end{itemize}
In practice, Cora, Citeseer and Pubmed are regarded as homophilic graphs, and Chameleon, Squirrel, Actor, Cornell and Texas are considered as heterophilic graphs. 
The statistics of datasets are demonstrated in table \ref{tab:dataset}.

\section{Model Settings}
\label{app:setting}
For fair comparison, we reproduce all the baselines in our environment. For MLP, GCN, GAT and SGC model, we tune them for the optimal parameters. For FAGCN, GPR-GNN, ACM-GCN, BM-GCN and GloGNN++, we rerun the models with the default parameters given by the author. For GeomGCN, H$_2$GCN, BernNet and HOC-GCN, we report the results of published papers.

For our method, we use early stopping strategy with 200 epochs and set an maximum epoch number as 1000. We set the dimension of node embedding and relation vector as 64, and the layer number of GNN as 2. We use Adam optimizer to train the model, and tune learning rate from $\{0.01,0.02,0.05,0.001,0.002,0.005\}$, weight decay from $\{0,1e-4,5e-4,1e-5,5e-5\}$. We tune the hyperparameters $\alpha_s, \alpha_f, \alpha_l, \theta, \gamma$ from 0 to 1, with 0.1 step size. To mitigate the overfitting problem, we take dropout when training. We implement our method based on PyTorch Geometric (PyG) library \cite{pyg} and Python 3.9.12. The program is executed on 32GB Tesla V100 GPU.
\section{Visualization Analysis of Relation Vectors}
\label{app:rel}
\begin{figure*}[ht]
	\centering
	\subfigure[Signed Message]{
		\includegraphics[width= \columnwidth]{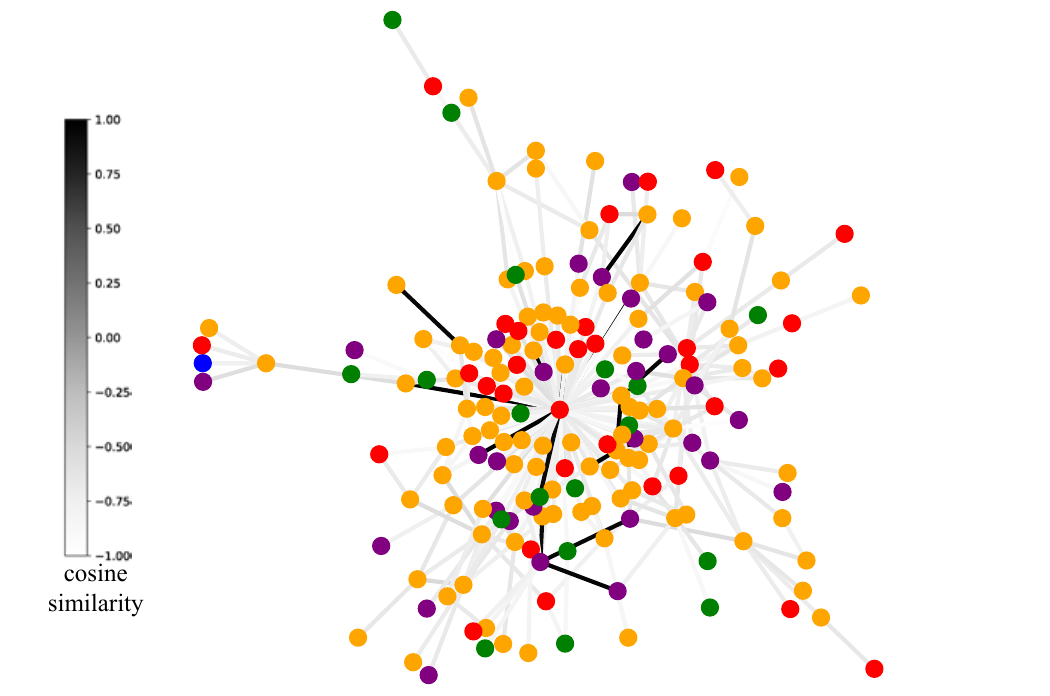}
	}
	\subfigure[Relation Vector Message]{
		\includegraphics[width= \columnwidth]{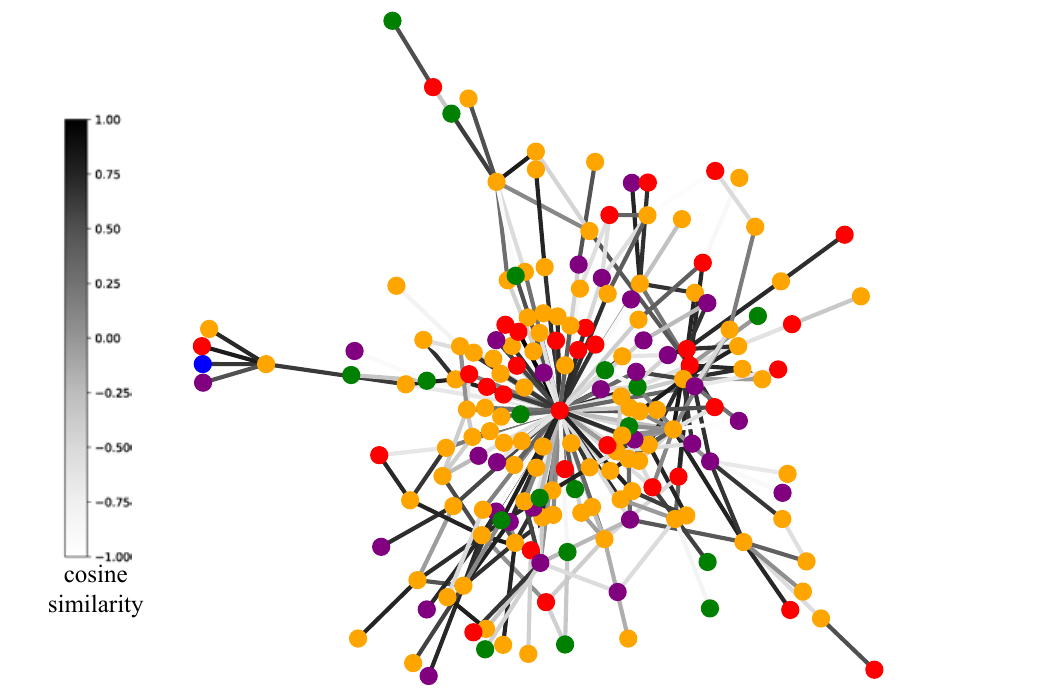}
	}
	\caption{Similarity visualization between the message passed and the central node class. 
    Different node colors represent different classes. 
    The edge shade indicates the cosine similarity value from $-1$ to $1$.}
	\label{fig:rel_an}
\end{figure*}
\begin{figure}[ht]
	\centering
	\subfigure[Homophily Connection]{
		\includegraphics[width=0.47 \columnwidth]{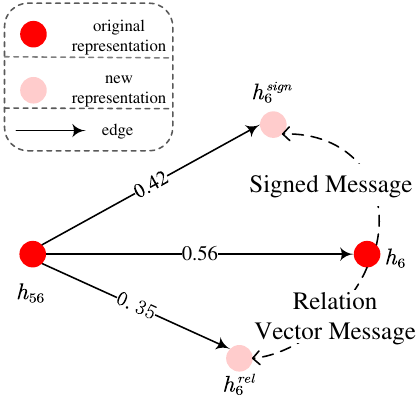}
	}
	\subfigure[Heterophily Connection]{
		\includegraphics[width=0.47 \columnwidth]{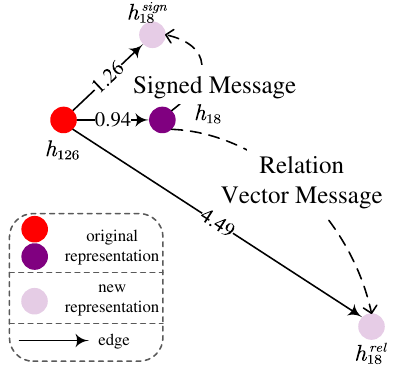}
	}
	\caption{
  Relation vector message and signed message for assimilating and dissimilating nodes. 
  The two edges are $e_{56, 6}$ and $e_{126,18}$ in Texas dataset.
  Red and purple colors denote the node classes, and light coloring represents the new representation after ``aggregation''. 
  The number of edges indicate the Euclidean distance of two nodes.}
	\label{fig:rel_ca}
\end{figure}

In this section, we give some intuitive demonstrations of the relation vector and compare it with the signed message method.

For the message passing on edge $e_{ji}$, we denote the relation vector message as $\mb{m}_{ji}^{rel}$ and compute it according to equation \ref{eq:agg}: 
\begin{equation*}
	\begin{split}
    \mb{m}_{ji}^{rel} = \mb{W}\mb{h}_{j}+\mb{z}_{ji}
	\end{split}
\end{equation*}
For the signed message, we directly assign the correct sign for the connection, with $+1$ for homophily and $-1$ for heterophily:
\begin{equation*}
	\begin{split}
    \mb{m}_{ji}^{sign}=
      \begin{cases}
        \mb{W}\mb{h}_{j}, & y_j = y_i \\
        - \mb{W}\mb{h}_{j}, & y_j \neq y_i
      \end{cases}
	\end{split}
\end{equation*}
where matrix $\mb{W}$ is the same in $\mb{m}_{ji}^{rel}$ and $\mb{m}_{ji}^{sign}$. Then we conduct two aspects of analysis, to compare the effect of two messages for modeling homophily/heterophily connections and helping node classification task.

\subsection{Effect for Modeling Homophily/Heterophily}
Modeling homophily/heterophily property of a connection means that the message could conduct effective assimilating/dissimilating operation between connected nodes.  
To evaluate this, we utilize the embedding of VR-GNN in $L-1$ layer and compute the relation vector message $\mb{m}_{ji}^{rel}$ and signed message $\mb{m}_{ji}^{sign}$ for each edge. Then by adding the messages to $\mb{h}_{i}$ respectively, we can get the new representation of central node $\mb{h}_{i}^{rel}$ and $\mb{h}_{i}^{sign}$. We compare $\mb{h}_{i}^{rel}$, $\mb{h}_{i}^{sign}$ and original representation $\mb{h}_{i}$ with the neighbor $\mb{h}_{j}$, to evaluate the distance change. 
For demonstration, we take two edges of Texas dataset as the example. The results are shown in figure \ref{fig:rel_ca}.
We can see that the ``aggregation'' of relation vector message makes the connected nodes closer under homophily and more distant under heterophily, which shows its superiority over signed message method. 

\subsection{Effect for Helping Node Classification}
In addition to accurately describing the connection property, a valid message should also be consistent with the central node class, so as to facilitate the node classification.
To show this, we first calculate the mean center for each node class, based on $L$ layer's node embedding of VR-GNN. 
Then we compute the message $\mb{m}_{ji}^{rel}$ and $\mb{m}_{ji}^{sign}$ of $L-1$ layer. 
For each edge, we compare $\mb{m}_{ji}^{rel}$ and $\mb{m}_{ji}^{sign}$ with the class center of node $v_i$ using cosine similarity. The results are shown in figure \ref{fig:rel_an}. 
We can observe that relation vector messages are more similar with the central node class than signed messages on most edges, which achieves a more center-cohesive node representation for classification and corresponds to the node visualization results in figure \ref{fig:visual}.  


\begin{table}[ht]
  \centering
  \begin{tabular}{l|cccc}
  \toprule[1pt]
  \multicolumn{1}{c|}{}                          & \multicolumn{4}{c}{Datasets}                                                                                                                   \\
  \multicolumn{1}{c|}{\multirow{-2}{*}{Encoder}} & Chameleon                         & Squirrel                          & Texas                             & Citeseer                          \\
  \midrule
  GCN                                            & {\color[HTML]{000000} 69.38} & {\color[HTML]{000000} 52.76} & {\color[HTML]{000000} 91.35} & {\color[HTML]{000000} 80.68} \\
  SGC                                            & {\color[HTML]{000000} 69.93} & {\color[HTML]{000000} 53.75} & {\color[HTML]{000000} 91.08} & {\color[HTML]{000000} 80.78} \\
  Ours                                            & {\color[HTML]{000000} \textbf{71.21}} & {\color[HTML]{000000} \textbf{57.50}} & {\color[HTML]{000000} \textbf{94.86}} & {\color[HTML]{000000} \textbf{81.95}} \\ \bottomrule[1pt] 
  \end{tabular}
  \caption{Comparison results with other encoder designs.}
  \label{tab:en}
\end{table}
\section{Comparison with Other Encoder Designs}
Currently, many works use GNN as an encoder to get the hidden embedding of data \cite{gvae,RI}.
To further verify the efficacy of our encoder design, we compare it with some GNN-based encoders such as GCN and SGC on four datasets. 
Specifically, we first employ GCN (SGC) to get node embeddings, then concatenate the embeddings of two endpoints of each edge, and use MLP to calculate the mean $\bm{\mu}$ and variance $\bm{\sigma}$ of the relation vector. Other structure of VR-GNN remains the same. The results are shown in table \ref{tab:en}.
We can observe that our encoder outperforms GCN and SGC by $3.1\%$ and $2.5\%$ on average. 
Our method is designed to explicitly extract three aspects of information for generation, which can make the encoding process more efficient and less noisy. 
Unlike GCN and SGC, we design to explicitly extracts three aspects of information for generation, which can make the encoding process more efficient and less noisy.
\end{document}